\documentclass[journal]{IEEEtran}

\usepackage[utf8]{inputenc} \usepackage[dvipsnames]{xcolor}
\usepackage[caption=false]{subfig} \usepackage{multirow}
\usepackage{graphicx} \usepackage{amsmath} \usepackage{listings}
\usepackage[cmintegrals]{newtxmath} \usepackage[free-standing-units,
per-mode=symbol,mode=text,binary-units=true,detect-weight=true,
detect-family=true]{siunitx} \usepackage{pgfplots} \usepackage{tikz}
\usepackage{paralist} \usepackage{booktabs} \usepackage{glossaries}
\usepackage[activate={true,nocompatibility},kerning=true,tracking=true,
spacing=true,factor=1100,stretch=20,shrink=20]{microtype}
\usepackage{algorithmic} \usepackage{textcomp} \usepackage{zi4}

\SetTracking{encoding={*}, shape=sc}{30}

\newcommand\apxref[1]{Appendix~\ref{#1}}
\newcommand\BigO[1]{\ensuremath{{\text{O}}{#1}}}
\newcommand\eg{e.g.,\ }
\newcommand\figref[1]{Figure~\ref{#1}}
\newcommand\ie{i.e.,\ }
\newcommand\etal{et~al.\ }
\newcommand\lstref[1]{Listing~\ref{#1}}
\newcommand\secref[1]{Section~\ref{#1}}
\newcommand\tabref[1]{Table~\ref{#1}}

\newcommand\review[1]{#1}

\DeclareSIUnit\bank{bank}
\DeclareSIUnit\cycle{cycle}
\DeclareSIPrefix\double{DP-}{0}
\DeclareSIPrefix\dpmega{DP-M}{6}
\DeclareSIPrefix\dpgiga{DP-G}{9}
\DeclareSIPrefix\dptera{DP-T}{12}
\DeclareSIUnit\flop{FLOP}
\DeclareSIUnit\flops{FLOPS}
\DeclareSIUnit\gate{GE}
\DeclareSIUnit\lane{lane}
\DeclareSIUnit\op{OP}
\DeclareSIUnit\ops{OPS}

\newacronym{ILP}{ILP}{Instruction Level Parallelism}
\newacronym{TLP}{TLP}{Thread Level Parallelism}
\newacronym{DLP}{DLP}{Data Level Parallelism}
\newacronym{ASIC}{ASIC}{Application-Specific Integrated Circuit}
\newacronym{AXI}{AXI}{Advanced eXtensible Interface}
\newacronym{MMU}{MMU}{Memory Management Unit}
\newacronym{FMA}{FMA}{Fused Multiply-Add}
\newacronym{RTL}{RTL}{Register Transfer Level}
\newacronym{FPU}{FPU}{Floating Point Unit}
\newacronym{ALU}{ALU}{Arithmetic Logic Unit}
\newacronym{MUL}{MUL}{multiplier}
\newacronym{SLDU}{SLDU}{Slide Unit}
\newacronym{PC}{PC}{Program Counter}
\newacronym{ROB}{ROB}{Re-order Buffer}
\newacronym{CNN}{CNN}{Convolutional Neural Network}
\newacronym{VNB}{VNB}{Von Neumann Bottleneck}
\newacronym{GPGPU}{GPGPU}{General-Purpose \acrlong{GPU}}
\newacronym{GPU}{GPU}{Graphics Processing Unit}
\newacronym{ECL}{ECL}{Emitter-Coupled Logic}
\newacronym{CMOS}{CMOS}{Complementary Metal-Oxide-Semiconductor}
\newacronym{FDSOI}{FD-SOI}{Fully Depleted Silicon on Insulator}
\newacronym{BLAS}{BLAS}{Basic Linear Algebra Subprograms}
\newacronym{VLIW}{VLIW}{Very Long Instruction Word}
\newacronym{CTS}{CTS}{Clock Tree Synthesis}
\newacronym{HDL}{HDL}{Hardware Description Language}
\newacronym{ISA}{ISA}{Instruction Set Architecture}
\newacronym{IPC}{IPC}{Instructions Per Cycle}
\newacronym{SM}{SM}{Streaming Multiprocessor}
\newacronym{PULP}{PULP}{Parallel Ultra Low Power}
\newacronym{BB}{BB}{Body-Biasing}
\newacronym{PG}{PG}{Power Gating}
\newacronym{RBB}{RBB}{Reverse Body-Biasing}
\newacronym{FBB}{FBB}{Forward Body-Biasing}
\newacronym{CSR}{CSR}{Control and State Register}
\newacronym{RVT}{RVT}{Regular Voltage Threshold}
\newacronym{LVT}{LVT}{Low Voltage Threshold}
\newacronym{SLVT}{SLVT}{Super Low Voltage Threshold}
\newacronym{DSP}{DSP}{Digital Signal Processing}
\newacronym{RoCC}{RoCC}{Rocket Custom Coprocessor Interface}
\newacronym{FPGA}{FPGA}{Field-Programmable Gate Array}
\newacronym{MIMD}{MIMD}{multiple instruction, multiple data}
\newacronym{SIMD}{SIMD}{single instruction, multiple data}
\newacronym{SIMT}{SIMT}{single instruction, multiple thread}
\newacronym{VT}{VT}{vector thread}
\newacronym{VLR}{VLR}{vector length register}
\newacronym{MVL}{MVL}{maximum vector length}
\newacronym{VRF}{VRF}{Vector Register File}
\newacronym{VFU}{VFU}{vector functional unit}
\newacronym{PTW}{PTW}{page-table walker}
\newacronym{PTE}{PTE}{page-table entry}
\newacronym{VID}{VID}{Vector Instruction Decode}
\newacronym{VIS}{VISSUE}{Vector Instruction Issue}
\newacronym{VLOOP}{VLOOP}{Vector Loop}
\newacronym{VAC}{VAC}{Vector Access}
\newacronym{VCONV}{VCONV}{Vector Conversion}
\newacronym{VEX}{VEX}{Vector Execute}
\newacronym{VLSU}{VLSU}{Vector Load/Store Unit}
\newacronym{LSU}{LSU}{Load/Store Unit}
\newacronym{RAW}{RAW}{read-after-write}
\newacronym{WAR}{WAR}{write-after-read}
\newacronym{WAW}{WAW}{write-after-write}
\newacronym{SSE}{SSE}{Streaming SIMD Extension}
\newacronym{AVX}{AVX}{Advanced Vector Extension}
\newacronym{SVE}{SVE}{Scalable Vector Extension}
\newacronym{HPC}{HPC}{High-Performance Computing}
\newacronym{PE}{PE}{processing element}

\colorlet{color1}{MidnightBlue}
\colorlet{color2}{PineGreen}
\colorlet{color3}{Plum}
\colorlet{color4}{YellowOrange}
\colorlet{colorAlert}{Red}

\lstdefinelanguage{rv}
{
  keywords={
    add,
    sub,
    vld,
    ld,
    vins,
    vmadd
  },
  sensitive=false,
  morecomment=[l]{;}
}

\makeatletter
\def\lst@makecaption{%
  \def\@captype{table}%
  \@makecaption
}
\makeatother

\usetikzlibrary{scopes, calc, shapes, arrows, positioning}
\tikzset{>=latex}

\pgfplotsset{compat=1.13}
\pgfplotsset{width=\linewidth, height=7cm}
\pgfplotsset{every x tick label/.append style={font=\small}}
\pgfplotsset{every y tick label/.append style={font=\small}}

\pgfplotsset{
  /pgf/declare function={
    roof(\x,\b,\p) = (\b * \x < \p) * \b * \x + (\b * \x >= \p) * \p;}}

\usepgfplotslibrary{external}
\usepgfplotslibrary{groupplots}
\usepgfplotslibrary{fillbetween}

\begin{document}


\title{Ara: A \SI{1}{\giga\hertz}+ Scalable and Energy-Efficient
  RISC-V Vector Processor with Multi-Precision Floating Point Support
  in \SI{22}{\nano\meter} FD-SOI}

\author{Matheus~Cavalcante,$^*$ Fabian~Schuiki,$^*$
  Florian~Zaruba,$^*$ Michael~Schaffner,$^*$
  Luca~Benini,$^{*\kern-0.05em\dagger}$~\IEEEmembership{Fellow,~IEEE}
  \thanks{$^*$\kern-0.1emIntegrated Systems Laboratory of ETH
    Z\"urich, Z\"urich, Switzerland. $^\dagger$\kern-0.1emDepartment
    of Electrical, Electronic, and Information Engineering Guglielmo
    Marconi of the University of Bologna, Bologna, Italy. E-mail:
    \{matheusd, fschuiki, zarubaf, mschaffner, lbenini\} \emph{at}
    iis.ee.ethz.ch.}}

\maketitle

\begin{abstract}
  In this paper, we present Ara, a 64-bit vector processor based on
  the version 0.5 draft of RISC-V's vector extension, implemented in
  \textsc{GlobalFoundries} 22FDX FD-SOI technology. Ara's
  microarchitecture is scalable, as it is composed of a set of
  identical lanes, each containing part of the processor's vector
  register file and functional units. It achieves up to
  \num{97}{\percent} FPU utilization when running a \num{256 x 256}
  double precision matrix multiplication on sixteen lanes. Ara runs at
  \review{more than \SI{1}{\giga\hertz}} in the typical corner
  (TT/\SI{0.80}{\volt}\kern-0.12em/\SI{25}{\celsius}), achieving a
  performance up to \SI{33}{\dpgiga\flops}. In terms of energy
  efficiency, Ara achieves up to \SI{41}{\dpgiga\flops\per\watt} under
  the same conditions, which is slightly superior to similar vector
  processors found in literature. An analysis on several vectorizable
  linear algebra computation kernels for a range of different matrix
  and vector sizes gives insight into performance limitations and
  bottlenecks for vector processors and outlines directions to
  maintain high energy efficiency even for small matrix sizes where
  the vector architecture achieves suboptimal utilization of the
  available FPUs.
\end{abstract}


\IEEEpeerreviewmaketitle

\glsresetall


\section{Introduction}
\label{sec:intro}

\IEEEPARstart{T}{he} end of Dennard scaling caused the race for
performance through higher frequencies to halt more than a decade ago,
when an increasing integration density stopped translating into
proportionate increases in performance or energy
efficiency~\cite{Dreslinski2010}. Processor frequencies plateaued,
inciting interest in parallel multi-core architectures. \review{These
  architectures, however, fail to address the efficiency limitation
  created by the inherent fetching and decoding of elementary
  instructions, which only keep the processor datapath busy for a very
  short period of time. Moreover, power dissipation limits how much
  integrated logic can be turned on simultaneously, increasing the
  energy efficiency requirements of modern
  systems~\cite{Hwang2016,Kiamehr2017}.}

In instruction-based programmable architectures, the key challenge is
how to mitigate the \gls{VNB}~\cite{Backus1978}. Despite the
flexibility of multi-core designs, they fail to exploit the regularity
of data-parallel applications. Each core tends to execute the same
instructions many times---a waste in terms of both area and
energy~\cite{Dabbelt2016}. The strong emergence of massively
data-parallel workloads, such as data analytics and machine
learning~\cite{Sze2017}, created a major window of opportunity for
architectures that effectively exploit data parallelism to achieve
energy efficiency. The most successful of these architectures are
General Purpose \glspl{GPU}~\cite{Owens2008}, which heavily leverage
data-parallel multithreading to relax the \gls{VNB} through the
so-called \gls{SIMT} approach~\cite{Lindholm2008}. \glspl{GPU}
dominate the energy efficiency race, being present in
\num{70}{\percent} of the Green500 ranks~\cite{Green5002018}. They are
also highly successful as data-parallel accelerators in
high-performance embedded applications, such as self-driving
cars~\cite{Bojarski206}.

The quest for extreme energy efficiency in data-parallel execution has
also revamped interest on vector architectures. This kind of
architecture was cutting-edge during another technology scaling
crisis, namely the one related to circuits based on the
Emitter-Coupled Logic technology~\cite{Russell1978}. Today, designers
and architects are reconsidering vector processing approaches, as they
promise to address the \gls{VNB} very effectively~\cite{Beldianu2015},
providing better energy efficiency than a general-purpose processor
for applications that fit the vector processing
model~\cite{Dabbelt2016}. A single vector instruction can be used to
express a data-parallel computation on a very large vector, thereby
amortizing the instruction fetch and decode overhead. The effect is
even more pronounced than for \gls{SIMT} architectures, where
instruction fetches are only amortized over the number of parallel
scalar execution units in a ``processing block'': for the latest
NVIDIA Volta \glspl{GPU}, such blocks are only 32 elements
long~\cite{Volta2017}. Therefore, vector processors provide a notably
effective model to efficiently execute the data parallelism of
scientific and matrix-oriented computations~\cite{Mano2015,
  Hennessy2011}, as well as digital signal processing and machine
learning algorithms.

The renewed interest in vector processing is reflected by the
introduction of vector instruction extensions in all popular
\glspl{ISA}, such as the proprietary ARM \gls{ISA}~\cite{Stephens2017}
and the open-source RISC-V \gls{ISA}~\cite{RVBase2019}. In this paper,
we set out to analyze the scalability and energy efficiency of vector
processors by designing and implementing a RISC-V-based architecture
in an advanced \gls{CMOS} technology. The design will be open-sourced
under a liberal license as part of the PULP Platform\footnote{See
  https://pulp-platform.org/.}\kern-.1em. The key contributions of
this paper are:
\begin{enumerate}
\item The architecture of a parametric in-order high-performance
  64-bit vector unit based on the version 0.5 draft of RISC-V's vector
  extension~\cite{RISCV2019}. The vector processor was designed for a
  memory bandwidth per peak performance ratio of
  \SI{2}{\byte\per\double\flop}, and works in tandem with Ariane, an
  open-source application-class RV64GC scalar core. The vector unit
  supports mixed-precision arithmetic with double, single, and
  half-precision floating point operands.
\item Performance analysis on key data-parallel kernels, both compute-
  and memory-bound, for variable problem sizes and design
  parameters. The performance is shown to meet the roofline achievable
  performance boundary, as long as the vector length is at least a few
  times longer than the number of physical lanes.
\item An architectural exploration and scalability analysis of the
  vector processor with post-implementation results extracted from
  \textsc{GlobalFoundries} 22FDX \gls{FDSOI} technology.
\item Insights on performance limitations and bottlenecks, for both
  the proposed architecture and for other vector processors found in
  the literature.
\end{enumerate}

This paper is organized as follows. In \secref{sec:backgr-relat-work}
we present some background and related work with the architectural
models most commonly used to explore data parallelism. Then, in
\secref{sec:archi}, we present the architecture of our vector
processor. \secref{sec:benchmarks} presents the benchmarks we used to
evaluate our vector unit. \secref{sec:performance-analysis} analyzes
how our vector unit explores \gls{HPC} workloads in terms of
performance, while \secref{sec:results} analyzes implementation
results in terms of power and energy efficiency. Finally,
\secref{sec:conclusions} concludes the paper and outlines future
research directions.


\section{Background and Related work}
\label{sec:backgr-relat-work}

\Gls{SIMD} architectures share---thus amortize---the instruction fetch
among multiple identical processing units. This architectural model
can be seen as instructions operating on vectors of operands. The
approach works well as long as the control flow is regular, \ie it is
possible to formulate the problem in terms of vector operations.

\subsection{Array processors}
\label{sec:packed-simd}

\begin{figure}[ht]
  \centering

  \begin{tikzpicture}[scale=.85,/tikz/font=\small]
    \draw (-1.3,2.7) rectangle node [midway] {\gls{PE}$_0$} ++(.8,.8);
    \draw (-1.3,1.8) rectangle node [midway] {\gls{PE}$_1$} ++(.8,.8);
    \draw (-1.3,0.9) rectangle node [midway] {\gls{PE}$_2$} ++(.8,.8);
    \draw (-1.3,0.0) rectangle node [midway] {\gls{PE}$_3$} ++(.8,.8);

    \draw[thick, ->] (-.08, -.2) -- ++(4.3,0) node [right] {$t$};

    \draw[rounded corners, fill=gray!30] (-0.05, -0.05) rectangle ++(.9,3.6);
    \draw[rounded corners, fill=gray!30] ( 0.95, -0.05) rectangle ++(.9,3.6);
    \draw[rounded corners, fill=gray!30] ( 1.95, -0.05) rectangle ++(.9,3.6);
    \draw[rounded corners, fill=gray!30] ( 2.95, -0.05) rectangle ++(.9,3.6);

    \draw[rounded corners, fill=white] (0,2.7) rectangle node [midway] {\emph{ld}$_0$} ++(.8,.8);
    \draw[rounded corners, fill=white] (0,1.8) rectangle node [midway] {\emph{ld}$_1$} ++(.8,.8);
    \draw[rounded corners, fill=white] (0,0.9) rectangle node [midway] {\emph{ld}$_2$} ++(.8,.8);
    \draw[rounded corners, fill=white] (0,0.0) rectangle node [midway] {\emph{ld}$_3$} ++(.8,.8);

    \draw[rounded corners, fill=white] (1,2.7) rectangle node [midway] {\emph{mul}$_0$} ++(.8,.8);
    \draw[rounded corners, fill=white] (1,1.8) rectangle node [midway] {\emph{mul}$_1$} ++(.8,.8);
    \draw[rounded corners, fill=white] (1,0.9) rectangle node [midway] {\emph{mul}$_2$} ++(.8,.8);
    \draw[rounded corners, fill=white] (1,0.0) rectangle node [midway] {\emph{mul}$_3$} ++(.8,.8);

    \draw[rounded corners, fill=white] (2,2.7) rectangle node [midway] {\emph{add}$_0$} ++(.8,.8);
    \draw[rounded corners, fill=white] (2,1.8) rectangle node [midway] {\emph{add}$_1$} ++(.8,.8);
    \draw[rounded corners, fill=white] (2,0.9) rectangle node [midway] {\emph{add}$_2$} ++(.8,.8);
    \draw[rounded corners, fill=white] (2,0.0) rectangle node [midway] {\emph{add}$_3$} ++(.8,.8);

    \draw[rounded corners, fill=white] (3,2.7) rectangle node [midway] {\emph{st}$_0$} ++(.8,.8);
    \draw[rounded corners, fill=white] (3,1.8) rectangle node [midway] {\emph{st}$_1$} ++(.8,.8);
    \draw[rounded corners, fill=white] (3,0.9) rectangle node [midway] {\emph{st}$_2$} ++(.8,.8);
    \draw[rounded corners, fill=white] (3,0.0) rectangle node [midway] {\emph{st}$_3$} ++(.8,.8);
  \end{tikzpicture}

  \caption{Execution pattern on an array processor~\cite{Flynn1972}.}
  \label{fig:arrayprocessor}
\end{figure}
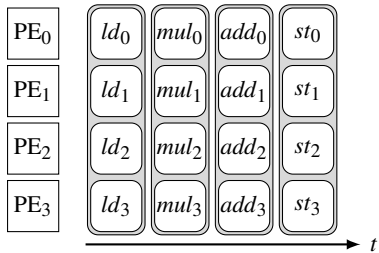

Array processors implement a packed-\gls{SIMD} architecture. This type
of processor has several independent but identical \glspl{PE}, all
operating on commands from a shared control unit.
\figref{fig:arrayprocessor} shows an execution pattern for a dummy
instruction sequence. The number of \glspl{PE} determines the vector
length, and the architecture can be seen as a wide datapath
encompassing all subwords, each handled by a
\gls{PE}~\cite{Peleg1996}.

A limitation of such an architecture is that the vector length is
fixed. It is commonly encoded into the instruction itself, meaning
that each expansion of the vector length comes with another \gls{ISA}
extension. For instance, Intel's first version of the \glspl{SSE}
operates on \SI{128}{\bit} registers, whereas the \gls{AVX} and
\gls{AVX}-512 evolution operates on \num{256} and \num{512}-{\bit}
wide registers, respectively~\cite{Reinders2017}. ARM provides
packed-\gls{SIMD} capability via the ``Neon'' extension, operating on
\SI{128}{\bit} wide registers~\cite{ArmNeon2019}. RISC-V also supports
packed-\gls{SIMD} via DSP extensions~\cite{Gautschi2017}.

\subsection{Vector processors}
\label{sec:vector-simd}

\begin{figure}[ht]
  \centering

  \begin{tikzpicture}[scale=.85,/tikz/font=\small]
    \draw (-1.3,2.85) rectangle node [midway] {\textsc{ld}} ++(.8,.8);
    \draw (-1.3,1.90) rectangle node [midway] {\textsc{mul}} ++(.8,.8);
    \draw (-1.3,0.95) rectangle node [midway] {\textsc{alu}} ++(.8,.8);
    \draw (-1.3,0.00) rectangle node [midway] {\textsc{st}} ++(.8,.8);

    \draw[thick, ->] (-.08, -.2) -- ++(7.3,0) node [right] {$t$};

    \draw[rounded corners, fill=gray!30] (-0.05,  2.80) rectangle ++(3.9,.9);
    \draw[rounded corners, fill=gray!30] ( 0.95,  1.85) rectangle ++(3.9,.9);
    \draw[rounded corners, fill=gray!30] ( 1.95,  0.90) rectangle ++(3.9,.9);
    \draw[rounded corners, fill=gray!30] ( 2.95, -0.05) rectangle ++(3.9,.9);

    \draw[rounded corners, fill=white] (0,2.85) rectangle node [midway] {\emph{ld}$_0$} ++(.8,.8);
    \draw[rounded corners, fill=white] (1,2.85) rectangle node [midway] {\emph{ld}$_1$} ++(.8,.8);
    \draw[rounded corners, fill=white] (2,2.85) rectangle node [midway] {\emph{ld}$_2$} ++(.8,.8);
    \draw[rounded corners, fill=white] (3,2.85) rectangle node [midway] {\emph{ld}$_3$} ++(.8,.8);

    \draw[rounded corners, fill=white] (1,1.90) rectangle node [midway] {\emph{mul}$_0$} ++(.8,.8);
    \draw[rounded corners, fill=white] (2,1.90) rectangle node [midway] {\emph{mul}$_1$} ++(.8,.8);
    \draw[rounded corners, fill=white] (3,1.90) rectangle node [midway] {\emph{mul}$_2$} ++(.8,.8);
    \draw[rounded corners, fill=white] (4,1.90) rectangle node [midway] {\emph{mul}$_3$} ++(.8,.8);

    \draw[rounded corners, fill=white] (2,0.95) rectangle node [midway] {\emph{add}$_0$} ++(.8,.8);
    \draw[rounded corners, fill=white] (3,0.95) rectangle node [midway] {\emph{add}$_1$} ++(.8,.8);
    \draw[rounded corners, fill=white] (4,0.95) rectangle node [midway] {\emph{add}$_2$} ++(.8,.8);
    \draw[rounded corners, fill=white] (5,0.95) rectangle node [midway] {\emph{add}$_3$} ++(.8,.8);

    \draw[rounded corners, fill=white] (3,0) rectangle node [midway] {\emph{st}$_0$} ++(.8,.8);
    \draw[rounded corners, fill=white] (4,0) rectangle node [midway] {\emph{st}$_1$} ++(.8,.8);
    \draw[rounded corners, fill=white] (5,0) rectangle node [midway] {\emph{st}$_2$} ++(.8,.8);
    \draw[rounded corners, fill=white] (6,0) rectangle node [midway] {\emph{st}$_3$} ++(.8,.8);
  \end{tikzpicture}

  \caption{Execution pattern on a vector processor~\cite{Flynn1972}.}
  \label{fig:vectorprocessor}
\end{figure}
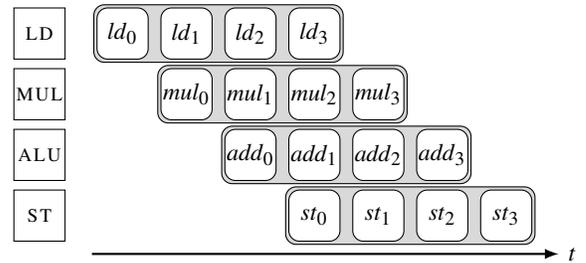

Vector processors are time-multiplexed versions of array processors,
implementing vector-\gls{SIMD} instructions. Several specialized
functional units stream the micro-operations on consecutive cycles, as
shown in \figref{fig:vectorprocessor}. By doing so, the number of
functional units no longer constrains the vector length, which can be
dynamically configured. As opposed to packed-\gls{SIMD}, long vectors
do not need to be subdivided into fixed-size chunks, but can be issued
using a single vector instruction. Hence, vector processors are
potentially more energy efficient than an equivalent array processor
since many control signals can be kept constant throughout the
computation, and the instruction fetch cost is amortized among many
cycles.

The history of vector processing starts with the traditional vector
machines from the sixties and seventies, with the beginnings of the
Illiac IV project~\cite{Mano2015}. The trend continued throughout the
next two decades, with work on supercomputers such as the
Cray-1~\cite{Russell1978}. At the end of the century, however,
microprocessor-based systems approached or surpassed the performance
of vector supercomputers at much lower costs~\cite{Asanovic1998}, due
to intense work on superscalar and \gls{VLIW} architectures. It is
only recently that vector processors got renewed interest from the
scientific community.

\review{Vector processors found a way into \glspl{FPGA} as
  general-purpose accelerators. VIPERS~\cite{Yu2009} is a vector
  processor architecture loosely compliant with
  VIRAM~\cite{Kozyrakis2003b}, with several \gls{FPGA}-specific
  optimizations. VEGAS~\cite{Chou2011} is a soft vector processor
  operating directly on scratchpad memory instead of on a \gls{VRF}.}

ARM is moving into Cray-inspired processing with their
\gls{SVE}~\cite{Stephens2017}. The extension is based on the vector
register architecture introduced with the Cray-1, and leaves the
vector length as an implementation choice (from \SI{128}{\bit} to
\SI{2048}{\bit}, in \SI{128}{\bit} increments). It is possible to
write code agnostic to the vector length, so that different
implementations can run the same software. The first system to adopt
this extension is Fujitsu's A64FX, at a peak performance of
\SI{2.7}{\dptera\flops} in a \SI{7}{\nano\meter} process, which is
competitive in terms of peak performance to leading-edge
\glspl{GPU}~\cite{Yoshida2018}.

The open RISC-V \gls{ISA} specification is also leading an effort
towards vector processing through its vector
extension~\cite{RISCV2019}. This extension is in active development,
and, at the time of this writing, its latest version was the 0.7. When
compared with ARM \gls{SVE}, RISC-V does not put any limits on the
vector length. Moreover, the extension makes it possible to trade off
the number of architectural vector registers against longer
vectors. Due to the availability of open-source RISC-V scalar cores,
together with the liberal license of the \gls{ISA} itself, we chose to
design our vector processor based on this extension.

One crucial issue in vector processing design is how to maximize the
utilization of the vector lanes. \review{Beldianu and
  Ziavras~\cite{Beldianu2015} and Lu~\etal\cite{Lu2016} explore
  sharing a pool of vector units among different threads. The
  intelligent sharing of vector units on multi-core increases their
  efficiency and throughput when compared to multi-core with per-core
  private vector units~\cite{Beldianu2015}. A \num{32}-bit
  implementation of the idea at \textsc{TSMC} \SI{40}{\nm} process is
  presented at~\cite{Beldianu2014}. However, the \gls{ISA} considered
  at such implementation is limited~\cite{Lu2016} when compared to
  RISC-V's vector extension, lacking, for example, the \gls{FMA}
  instruction, strictly required in high-performance
  workloads. Moreover, the wider \num{64}-bit datapath of our vector
  unit implies a drastic complexity increase of the \gls{FMA} units
  and a larger \gls{VRF}, and consequently a quantitative energy
  efficiency comparison between Ara and~\cite{Beldianu2014} is not
  directly possible. We compared the achieved vector lanes'
  utilization in \secref{sec:matr-mult}.}

\subsection{SIMT}
\label{sec:simt}

\gls{SIMT} architectures represent an amalgamation of the flexibility
of \gls{MIMD} and the efficiency of \gls{SIMD} designs. While
\gls{SIMD} architectures apply one instruction to multiple data lanes,
\gls{SIMT} designs apply one instruction to multiple independent
threads in parallel~\cite{Lindholm2008}. The NVIDIA Volta GV100
\gls{GPU} is a state-of-the-art example of this architecture, with 64
``processing blocks,'' called \glspl{SM} by NVIDIA, each handling 32
threads.

A \gls{SIMD} instruction exposes the vector length to the programmer
and requires manual branching control, usually by setting flags that
indicate which lanes are active for a given vector
instruction. \gls{SIMT} designs, on the other hand, allow the threads
to diverge, although substantial performance improvement can be
achieved if they remain synchronized~\cite{Lindholm2008}. \gls{SIMD}
and \gls{SIMT} designs also handle data accesses differently. Since
\glspl{GPU} lack a control processor, hardware is necessary to
dynamically coalesce memory accesses into large contiguous
chunks~\cite{Beldianu2015}. While this approach simplifies the
programming model, it also incurs into a considerable energy
overhead~\cite{Lee2011}.

\subsection{Vector thread}
\label{sec:vector-thread}

Another compromise between \gls{SIMD} and \gls{MIMD} are \gls{VT}
architectures~\cite{Lee2011}, \review{which support loops with
  cross-iteration dependencies and arbitrary internal control
  flow~\cite{Krashinsky2004}}. Similar to \gls{SIMT} designs---and
unlike \gls{SIMD}---\gls{VT} architectures leverage the threading
concept instead of the more rigid notion of lanes, and hence provide a
mechanism to handle program divergence. The main difference between
\gls{SIMT} and \gls{VT} is that in the latter the vector instructions
reside in another thread, and scalar bookkeeping instructions can
potentially run concurrently with the vector ones. This division
alleviates the problem of \gls{SIMT} threads running redundant scalar
instructions that must be later coalesced in hardware. Hwacha is a
\gls{VT} architecture based on a custom RISC-V extension, recently
achieving \SI{64}{\dpgiga\flops} in \textsc{ST} \SI{28}{\nano\meter}
\gls{FDSOI} technology~\cite{Schmidt2018}.

Many vector architectures report only full-system metrics of
performance and efficiency, such as memory hierarchy or main memory
controllers. This is the case of Fujitsu's
A64FX~\cite{Yoshida2018}. As our focus is on the core execution
engine, we will mainly compare our vector unit with Hwacha in
\secref{sec:perf-power-area}. Hwacha is an open-sourced design
architecture for which information about the internal organization is
available, allowing for a fair quantitative comparison on a single
processing engine.


\section{Architecture}
\label{sec:archi}

\begin{figure*}[t]
  \centering
  \begin{minipage}[t]{0.5\linewidth}
    \centering \subfloat[Block diagram of an Ara instance with $N$
    parallel lanes. Ara receives its commands from Ariane, a RV64GC
    scalar core. The vector unit has a main sequencer; $N$ parallel
    lanes; a \gls{SLDU}; and a \gls{VLSU}. The memory interface is $W$
    \si{\bit}
    wide.\label{fig:archi}]{\includegraphics[width=.9\linewidth]{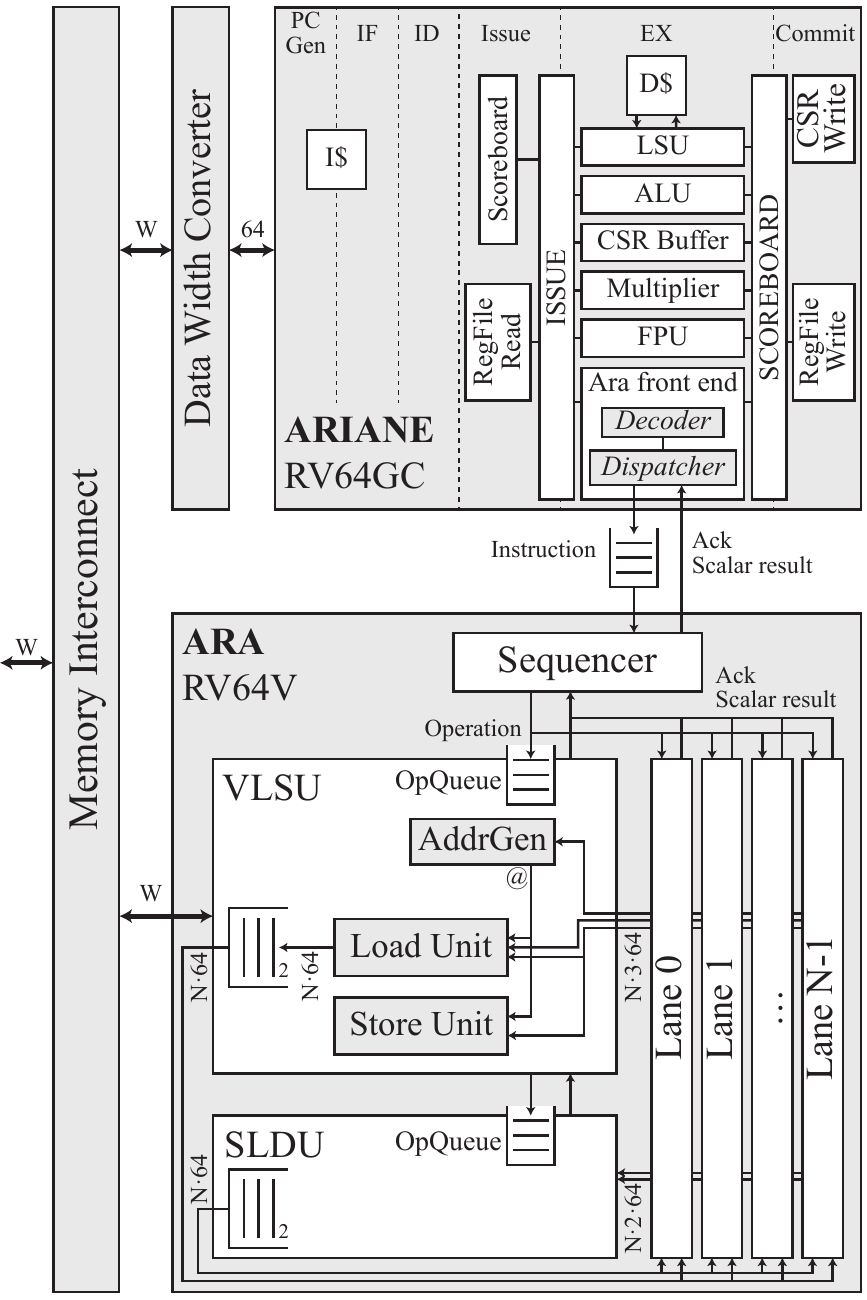}}
  \end{minipage}\hfill%
  \begin{minipage}[t]{0.5\linewidth}
    \centering \subfloat[Block diagram of one lane of Ara. It contains
    a lane sequencer (handling up to 8 vector instructions); a
    \SI{16}{\kibi\byte} vector register file; ten operand queues; an
    integer \gls{ALU}; an integer \gls{MUL}; and a
    \gls{FPU}.\label{fig:lane}]
    {\includegraphics[width=.9\linewidth]{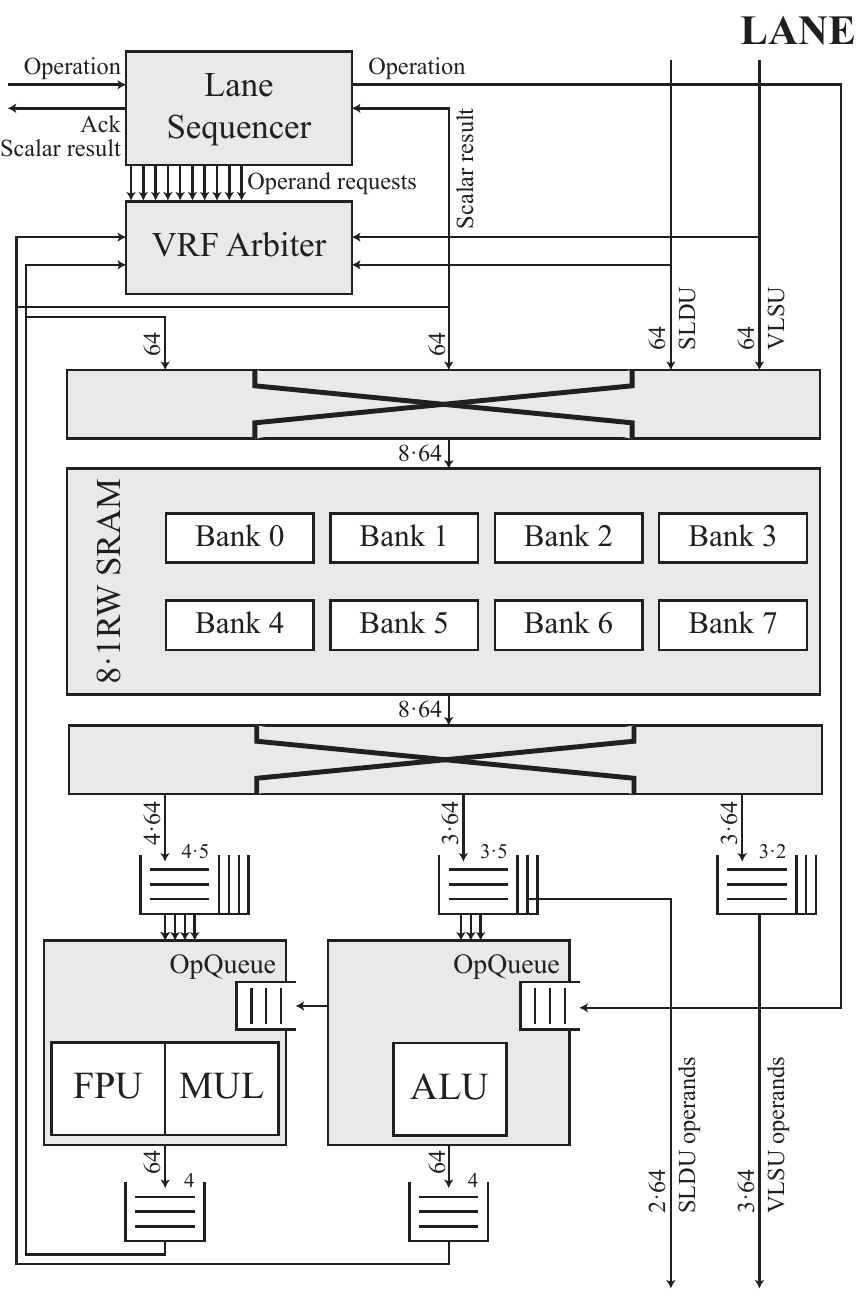}}
  \end{minipage}

  \caption{Top-level block diagram of Ara.}
  \label{fig:archi_top}
\end{figure*}

In this section, we introduce the microarchitecture of Ara, a scalable
high-performance vector unit based on RISC-V's vector extension. As
illustrated in \figref{fig:archi}, Ara works in tandem with
Ariane~\cite{Zaruba2019}, an open-source Linux-capable
application-class core. To this end, Ariane has been extended to drive
the accompanying vector unit as a tightly coupled coprocessor.

\subsection{Ariane}
\label{sec:ariane}

Ariane is an open-source, in-order, single-issue, 64-bit
application-class processor implementing RV64GC~\cite{Zaruba2019}. It
has support for hardware multiply/divide and atomic memory operations,
as well as an IEEE-compliant \gls{FPU}~\cite{Mach2018}. It has been
manufactured in \textsc{GlobalFoundries} 22FDX \gls{FDSOI} technology,
running at most at \SI{1.7}{\giga\hertz} and achieving an energy
efficiency of up to \SI{40}{\giga\ops\per\watt}. Zaruba and
Benini~\cite{Zaruba2019} report that the core has a six-stage
pipeline, namely \gls{PC} Generation, Instruction Fetch, Instruction
Decode, Issue Stage, Execute Stage, and Commit Stage. We denote the
first two stages as Ariane's front end, responsible for the
instruction fetch interface, and the remaining four as its back end.

Ariane needs some architectural changes to drive our vector unit, all
of them in the back end. Vector instructions are decoded partially in
Ariane's Instruction Decoder, to recognize whether they are vector
instructions, and then completely in a dedicated Vector Instruction
Decoder inside Ara. The reason for this split decoding is the high
number of Vector Control and Status Registers---one for each of the 32
vector registers---that are taken into account before fully decoding
such instructions.

The dispatcher controls the interface between Ara and Ariane's
dedicated scoreboard port. \review{In Ariane, instructions can retire
  out-of-order from the functional units~\cite{Zaruba2019}, while Ara
  executes instructions non-speculatively. The dispatcher also works
  speculatively, but waits until a vector instruction reaches the top
  of the scoreboard (\ie it is no longer speculative) to push it} into
the instruction queue, together with the contents of any scalar
registers read by the vector instruction. Ara reads from this queue,
and then acknowledges the instruction \review{(if required, \eg the
  vector instruction produces a scalar result)} or propagates
potential exceptions back to Ariane's scoreboard.

Instructions are acknowledged as soon as Ara determines that they will
not throw any exceptions. This happens early in their execution,
usually after their decoding. Because vector instructions can run for
an extended number of cycles (as presented in
\figref{fig:vectorprocessor}), they may get acknowledged many cycles
before the end of their execution, potentially freeing the scalar
cores to continue execution of its instruction stream. \review{The
  decoupled execution works well, except when Ariane expects a result
  from Ara, \eg reading an element of a vector register.}

\review{The interface between Ariane and Ara is lightweight, being
  similar to the \gls{RoCC}, for use with the Rocket
  Chip~\cite{Asanovic2016}. The difference between them is that
  dispatcher pushes the decoded instruction to Ara, while \gls{RoCC}
  leaves the full decoding task to the coprocessor.}

\subsection{Sequencer}
\label{sec:sequencer}

The sequencer is responsible for keeping track of the vector
instructions running on Ara, dispatching them to the different
execution units and acknowledging them with Ariane. This unit is the
single block that has a global view of the instruction execution
progress across all lanes. \review{The sequencer can handle up to
  eight parallel instructions. This ensures Ara has instructions
  enqueued for execution, avoiding starvation due to the
  non-speculative dispatch policy of Ara's front end.}

Hazards among pending vector instructions are resolved by this
block. Structural hazards arise due to architectural decisions (\eg
shared paths between the \gls{ALU} and the \gls{SLDU}) or if a
functional unit is not able to accept yet another instruction due to
the limited capacity of its operation queue. The sequencer delays the
issue of vector instructions until the structural hazard has been
resolved (\ie the offending instruction completes).

The sequencer also stores information about which vector instruction
is accessing which vector register. This information is used to
determine data hazards between instructions. For example, if a vector
instruction tries to write to a vector register that is already being
written, the sequencer will flag the existence of a \gls{WAW} data
hazard between them. \Gls{RAW}, \gls{WAR} and \gls{WAW} hazards are
handled in the same manner. Unlike structural hazards, data hazards do
not need to stall the sequencer, as they are handled on a per-element
basis downstream.

\subsection{Slide unit}
\label{sec:slide-unit}

The \gls{SLDU} is responsible for handling instructions that must
access all \gls{VRF} banks at once. It handles, for example, the
insertion of an element into a vector, the extraction of an element
from a vector, vector shuffles, and vector slides
($v_d[i] \gets v_s[i \,+\, {\text{slide amount}}]$). This unit may
also be extended to support basic vector reductions, such as
vector-add and internal product. The support for vector reductions is
considered an optional feature in the current version of RISC-V's
vector extension~\cite{RISCV2019}. For simplicity, we decided not to
support them, taking into consideration that an $\BigO{(n)}$ vector
reduction can still be implemented as a sequence of $\BigO{(\log n)}$
vector slides and the corresponding arithmetic
instruction~\cite{Asanovic1998}.

\subsection{Vector load/store unit}
\label{sec:vect-loadst-unit}

Ara has a single memory port, whose width is chosen to keep the memory
bandwidth per peak performance ratio fixed at
\SI{2}{\byte\per\double\flop}. As illustrated in \figref{fig:archi},
Ara has an address generator, responsible for determining which memory
address will be accessed. This can either be
\begin{inparaenum}[i)]
\item unit-stride loads and stores, which access a contiguous chunk of
  memory;
\item constant-stride memory operations, which access memory addresses
  spaced with a fixed offset; and
\item scatters and gathers, which use a vector of offsets to allow
  general access patterns.
\end{inparaenum}
After address generation, the unit coalesces unit-stride memory
operations into burst requests, avoiding the need to request the
individual elements from memory. The burst start address and the burst
length are then sent to either the load or the store unit, both of
which are responsible for initiating data transfers through Ara's
\gls{AXI} interface.

\subsection{Lane organization}
\label{sec:lane-organization}

Ara can be configured with a variable number of identical lanes, each
one with the architecture shown in \figref{fig:lane}. Each lane has
its own lane sequencer, responsible for keeping track of up to eight
parallel vector instructions. Each lane also has a \gls{VRF} and an
accompanying arbiter to orchestrate its access, operand queues, an
integer \gls{ALU}, an integer \gls{MUL}, and an \gls{FPU}.

Each lane contains part of Ara's whole \gls{VRF} and execution
units. Hence, most of the computation is contained within one lane,
and instructions that need to access all the \gls{VRF} banks at once
(\eg instructions that execute at the \gls{VLSU} or at the \gls{SLDU})
use data interfaces between the lanes and the responsible computing
units. Each lane also has a command interface attached to the main
sequencer, through which the lanes indicate they finished the
execution of an instruction.

\subsubsection{Lane sequencer}
\label{sec:lane-sequencer}

The lane sequencer is responsible for issuing vector instructions to
the functional units, controlling their execution in the context of a
single lane. Unlike the main sequencer, the lane sequencers do not
store the state of the running instructions, avoiding data duplication
across lanes. They also initiate requests to read operands from the
\gls{VRF}. We generate up to ten independent requests to the \gls{VRF}
arbiter.

Operand fetch and result write-back are decoupled from
each. Starvation is avoided via a self-regulated process, through back
pressure due to unavailable operands. By throttling the operation
request rate, the lane sequencer indirectly limits the rate at which
results are produced. This is used to handle data hazards, by ensuring
that dependent instructions run at the same pace: if instruction $i$
depends on instruction $j$, the operands of instruction $i$ are
requested only if instruction $j$ produced results in the previous
cycle. There is no forwarding logic.

\subsubsection{Vector register file}
\label{sec:vector-register-file}

The \gls{VRF} is at the core of every vector processor. Because
several instructions can run in parallel, the register file must be
able to support enough throughput to supply the functional units with
operands and absorb their results. In RISC-V's vector extension, the
predicated multiply-add instruction is the worst case regarding
throughput, reading four operands to produce one result.

Due to the massive area and power overhead of multi-ported memory
cuts, which usually require custom transistor-level design, we opted
not to use a monolithic \gls{VRF} with several ports. Instead, Ara's
vector register file is composed of a set of single-ported (1RW)
banks. The width of each bank is constrained to the datapath width of
each lane, \ie \SI{64}{\bit}, to avoid subword selection
logic. Therefore, in steady state, five banks are accessed
simultaneously to sustain maximum throughput for the predicated
multiply-add instruction. Ara's register file has eight banks per
lane, providing some margin on the banking factor. This \gls{VRF}
structure (eight \num{64}-{\bit} wide 1RW banks) is replicated at each
lane, and all inter-lane communication is concentrated at the
\gls{VLSU} and \gls{SLDU}. \review{We used a high-performance memory
  cut to meet a target operating frequency of
  \SI{1}{\giga\hertz}. These memories, however, cannot be fully
  clock-gated. The cuts do consume less power in idle state, a NOP
  costing about \num{10}{\percent} of the power required by a write
  operation.}

A multi-banked \gls{VRF} raises the problem of banking conflicts,
which occur when several functional units need to access the same
bank. These are resolved dynamically with a weighted round-robin
arbiter per bank with two priority levels. Low-throughput
instructions, such as memory operations, are assigned a lower
priority.  By doing so, their irregular access pattern does not
disturb other concurrent high-throughput instructions (\eg
floating-point instructions).

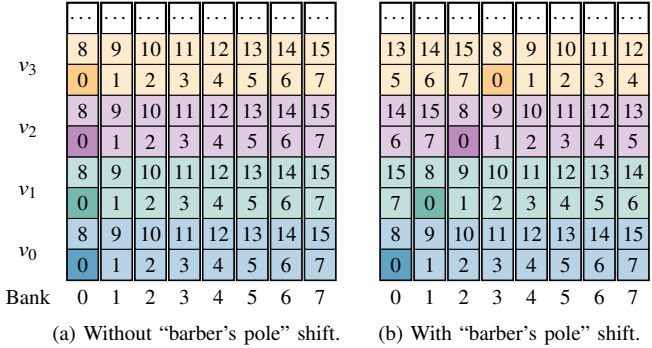
\begin{figure}[ht]
  \centering

  \begin{minipage}[t]{.06\linewidth}
    \centering
    \begin{tikzpicture}[/tikz/font=\footnotesize]
      \def\bwid{0.41}
      \def\bhei{0.41}

      \node at (0,0) {Bank};
      \node[align=center] at (0,1.4*\bhei) {$v_0$};
      \node[align=center] at (0,3.4*\bhei) {$v_1$};
      \node[align=center] at (0,5.4*\bhei) {$v_2$};
      \node[align=center] at (0,7.4*\bhei) {$v_3$};
    \end{tikzpicture}
  \end{minipage}\hfill%
  \begin{minipage}[t]{.46\linewidth}
    \centering
    \subfloat[Without ``barber's pole'' shift.\label{fig:vrf_organization_no_barber_pole}]{
      \begin{tikzpicture}[/tikz/font=\footnotesize]
        \def\bwid{0.41}
        \def\bhei{0.41}

        \foreach \b in {0,...,7} {
          \pgfmathsetmacro{\x}{1.1*\bwid*\b};
          \draw [thick] (\x,0) coordinate (A) rectangle ++(\bwid,9*\bhei) coordinate (B);
          \coordinate (M) at ($0.5*(A) + 0.5*(B)$);
          \node at (M |- A) [below] {\b};
        }

        \foreach \e in {0,...,15} {
          \pgfmathtruncatemacro{\b}{mod(\e,8)};
          \pgfmathtruncatemacro{\l}{\e / 8};
          \draw[fill=color1!20] (1.1*\bwid*\b, \bhei*\l) rectangle node[midway] {\num{\e}} ++(\bwid,\bhei);
        }
        \draw[fill=color1!50] (0, 0) rectangle node[midway] {$0$} ++(\bwid,\bhei);

        \foreach \e in {0,...,15} {
          \pgfmathtruncatemacro{\b}{mod(\e,8)};
          \pgfmathtruncatemacro{\l}{\e / 8 + 2};
          \draw[fill=color2!20] (1.1*\bwid*\b, \bhei*\l) rectangle node[midway] {\num{\e}} ++(\bwid,\bhei);
        }
        \draw[fill=color2!50] (0, 2*\bhei) rectangle node[midway] {$0$} ++(\bwid,\bhei);

        \foreach \e in {0,...,15} {
          \pgfmathtruncatemacro{\b}{mod(\e,8)};
          \pgfmathtruncatemacro{\l}{\e / 8 + 4};
          \draw[fill=color3!20] (1.1*\bwid*\b, \bhei*\l) rectangle node[midway] {\num{\e}} ++(\bwid,\bhei);
        }
        \draw[fill=color3!50] (0, 4*\bhei) rectangle node[midway] {$0$} ++(\bwid,\bhei);

        \foreach \e in {0,...,15} {
          \pgfmathtruncatemacro{\b}{mod(\e,8)};
          \pgfmathtruncatemacro{\l}{\e / 8 + 6};
          \draw[fill=color4!20] (1.1*\bwid*\b, \bhei*\l) rectangle node[midway] {\num{\e}} ++(\bwid,\bhei);
        }
        \draw[fill=color4!50] (0, 6*\bhei) rectangle node[midway] {$0$} ++(\bwid,\bhei);

        \foreach \b in {0,...,7} {
          \draw (1.1*\bwid*\b, \bhei*8) rectangle node[midway] {\dots} ++(\bwid,\bhei);
        }
      \end{tikzpicture}}
  \end{minipage}\hfill%
  \begin{minipage}[t]{.46\linewidth}
    \centering
    \subfloat[With ``barber's pole'' shift.\label{fig:vrf_organization_barber_pole}]{
      \begin{tikzpicture}[/tikz/font=\footnotesize]
        \def\bwid{0.41}
        \def\bhei{0.41}

        \foreach \b in {0,...,7} {
          \pgfmathsetmacro{\x}{1.1*\bwid*\b};
          \draw [thick] (\x,0) coordinate (A) rectangle ++(\bwid,9*\bhei) coordinate (B);
          \coordinate (M) at ($0.5*(A) + 0.5*(B)$);
          \node at (M |- A) [below] {\b};
        }

        \foreach \e in {0,...,15} {
          \pgfmathtruncatemacro{\b}{mod(\e,8)};
          \pgfmathtruncatemacro{\l}{\e / 8};
          \draw[fill=color1!20] (1.1*\bwid*\b, \bhei*\l) rectangle node[midway] {\num{\e}} ++(\bwid,\bhei);
        }
        \draw[fill=color1!50] (0, 0) rectangle node[midway] {$0$} ++(\bwid,\bhei);

        \foreach \e in {0,...,15} {
          \pgfmathtruncatemacro{\b}{mod(\e + 1,8)};
          \pgfmathtruncatemacro{\l}{\e / 8 + 2};
          \draw[fill=color2!20] (1.1*\bwid*\b, \bhei*\l) rectangle node[midway] {\num{\e}} ++(\bwid,\bhei);
        }
        \draw[fill=color2!50] (1.1*\bwid, 2*\bhei) rectangle node[midway] {$0$} ++(\bwid,\bhei);

        \foreach \e in {0,...,15} {
          \pgfmathtruncatemacro{\b}{mod(\e + 2,8)};
          \pgfmathtruncatemacro{\l}{\e / 8 + 4};
          \draw[fill=color3!20] (1.1*\bwid*\b, \bhei*\l) rectangle node[midway] {\num{\e}} ++(\bwid,\bhei);
        }
        \draw[fill=color3!50] (1.1*2*\bwid, 4*\bhei) rectangle node[midway] {$0$} ++(\bwid,\bhei);

        \foreach \e in {0,...,15} {
          \pgfmathtruncatemacro{\b}{mod(\e + 3,8)};
          \pgfmathtruncatemacro{\l}{\e / 8 + 6};
          \draw[fill=color4!20] (1.1*\bwid*\b, \bhei*\l) rectangle node[midway] {\num{\e}} ++(\bwid,\bhei);
        }
        \draw[fill=color4!50] (1.1*3*\bwid, 6*\bhei) rectangle node[midway] {$0$} ++(\bwid,\bhei);

        \foreach \b in {0,...,7} {
          \draw (1.1*\bwid*\b, \bhei*8) rectangle node[midway] {\dots} ++(\bwid,\bhei);
        }
      \end{tikzpicture}}
  \end{minipage}

  \caption{\gls{VRF} organization inside one lane. Darker colors
    highlight the initial element of each vector register $v_i$. In
    a), all vector registers start at the same bank. In b), the vector
    registers follow a ``barber's pole'' pattern, the starting bank
    being shifted for every vector register.}
  \label{fig:vrf_organization}
\end{figure}

\figref{fig:vrf_organization_barber_pole} shows how the vector
registers are mapped onto the banks. The initial bank of each vector
register is shifted in a ``barber's pole'' fashion. This avoids
initial banking conflicts when the functional units try to fetch the
first element of different vector registers, which are all mapped onto
the same bank in a pure element-partitioned
approach~\cite{Asanovic1998} of
\figref{fig:vrf_organization_no_barber_pole}.

\review{Vector registers can also hold scalar values. In this case,
  the scalar value is replicated at each lane at the first position of
  the vector register. Scalar values are only read/written once per
  lane, and are logically replicated by the functional units.}

\subsubsection{Operand queues}
\label{sec:operand-queues}

The multi-banked organization of the \gls{VRF} can lead to banking
conflicts when several functional units try to access operands in the
same bank. Each lane has a set of operand queues between the \gls{VRF}
and the functional units to absorb such banking conflicts. There are
ten operand queues: four of them are dedicated to the
\gls{FPU}/\gls{MUL} unit, three of them to the \gls{ALU} (two of which
are shared with the \gls{SLDU}), and another three to the
\gls{VLSU}. Each queue is \SI{64}{\bit} wide and their depth was
chosen via simulation. The queue depth depends on the functional
unit's latency and throughput, so that low-throughput functional
units, as the \gls{VLSU}, require shallower queues than the
\glspl{FPU}. Queues between the functional units' output ports and the
vector register file absorb banking conflicts on the write-back path
to the \gls{VRF}. Each lane has two of such queues, one for the
\gls{FPU}\kern-.1em/\gls{MUL} and one for the \gls{ALU}.
\review{Together with the decoupled operand fetch mechanism discussed
  in \secref{sec:lane-sequencer} and the barber's pole \gls{VRF}
  organization of \secref{sec:vector-register-file}, the operand
  queues allow for a pipelined execution of vector instructions. While
  bubbles occur sporadically due to banking conflicts, it is possible
  to fill the pipeline even with a succession of short vector
  instructions.}

\subsubsection{Execution units}
\label{sec:execution-units}

Each lane has three execution units, an integer \gls{ALU}, an integer
\gls{MUL}, and an \gls{FPU}, all of them operating on a \num{64}-bit
datapath. The \gls{MUL} shares the operand queues with the \gls{FPU},
and they cannot be used simultaneously, \review{since we do not expect
  the simultaneous use of the integer multiplier and the
  floating-point unit to be a common case}. With the exception of this
constraint, vector chaining is allowed between any execution units, as
long as they are executing instructions with regular access patterns
(\ie no vector shuffles).

It is possible to subdivide the \num{64}-bit datapath, trading off
narrower data formats by a corresponding increase in performance. The
three execution units have a \SI{64}{\bit\per\cycle} throughput,
regardless of the data format of the computation. We developed our
multi-precision \gls{ALU} and \gls{MUL}, both producing \num{1 x 64},
\num{2 x 32}, \num{4 x 16}, and \num{8 x 8}~{\bit} signed or unsigned
operands. Ara has limited support for multi-precision operations,
allowing for data promotions from \num{8} to \num{16}, \num{16} to
\num{32}, and from \num{32} to \SI{64}{\bit}.

For the \gls{FPU}, we used an open-source, IEEE-compliant,
multi-precision \gls{FPU} developed by Mach~\etal\cite{Mach2018}. The
\gls{FPU} was configured to support \glspl{FMA}, additions,
multiplications, divisions, square roots, and comparisons. As the
integer units, the \gls{FPU} has a \SI{64}{\bit\per\cycle} throughput,
\ie one double precision, two single precision or four IEEE 754
half-precision floating point results per cycle. Besides
IEEE~\num{754} standard floating point formats, the \gls{FPU} also
supports alternative formats, both \num{8}- and \num{16}-{\bit}
wide. Depending on the application, the narrower number formats can be
used to achieve significant energy savings compared to a wide
floating-point baseline~\cite{Mach2018}.


\section{Benchmarks}
\label{sec:benchmarks}

Memory bandwidth is often a limiting factor when it comes to processor
performance, and many optimizations revolve around scheduling memory
and arithmetic operations with the purpose of hiding memory
latency. The relationship between processor performance and memory
bandwidth can be analyzed with the roofline
model~\cite{Williams2009}. This model shows the peak achievable
performance (in \si{\op\per\cycle}) as a function of the arithmetic
intensity $I$\kern-.1em, defined as the algorithm-dependent ratio of
operations per byte of memory traffic.

Accordingly to this model, computations can be either memory-bound or
compute-bound~\cite{Ofenbeck2014}, the peak performance being
achievable only if the algorithm's arithmetic intensity, in operations
per byte, is higher than the processor's performance per memory
bandwidth ratio. For Ara, it enters its compute-bound regime when the
arithmetic intensity is higher than
\SI{0.5}{\double\flop\per\byte}. The memory bandwidth determines the
slope of the performance boundary in the memory-bound regime. We
consider three benchmarks to explore the architecture instances of the
vector processor with distinct arithmetic intensities that fully span
the two regions of the roofline.

Our first algorithm is MATMUL, a $n \times n$ double-precision matrix
multiplication $C \gets AB + C$. The algorithm requires $2n^3$
floating-point operations---one \gls{FMA} is considered as two
operations---and at least $32n^2$ bytes of memory
transfers. Therefore, the algorithm has an arithmetic intensity of at
least
\begin{equation}
  \label{eq:3}
  I_{{\text{MATMUL}}} \ge \frac{n}{16}~\si{\double\flop\per\byte}.
\end{equation}
We will consider matrices of size at least \num{16 x 16} across
several Ara instances. The roofline model shows that it is possible to
achieve the system's peak performance with these matrix sizes.

Matrix multiplication is neither embarrassingly memory-bound nor
compute-bound, since its arithmetic intensity grows with
$\BigO{(n)}$. Nevertheless, it is interesting to see how Ara behaves
on highly memory-bound as well as fully compute-bound cases. DAXPY,
$Y \gets \alpha X + Y$, is a common algorithmic building block of more
complex \gls{BLAS} routines. Considering vectors of length $n$, DAXPY
requires $n$ \glspl{FMA} and at least $24n$ bytes of memory
transfers. DAXPY is therefore a heavily memory-bound algorithm, with
an arithmetic intensity of \SI{1/12}{\double\flop\per\byte}.

We explore the extremely compute-bound spectrum with the tensor
convolution DCONV, a routine which is at the core of convolutional
networks. In terms of size, we took the first layer of
GoogLeNet~\cite{Szegedy2015}, with a \num{64 x 3 x 7 x 7} kernel and
\num{3 x 112 x 112} input images. Each point of the input image must
be convolved with the weights, resulting in a total of \num{64 x 3 x 7
  x 7 x 112 x 112} \glspl{FMA}, or \SI{236}{\dpmega\flop}. In terms of
memory, we will consider that the input matrix (after padding) is
loaded exactly once, or \num{3 x 118 x 118} double precision loads,
together with the write-back of the result, or \num{64 x 112 x 112}
double precision stores. The \SI{6.44}{\mebi\byte} of memory transfers
imply an arithmetic intensity of \SI{34.9}{\double\flop\per\byte},
making this kernel heavily compute-bound on Ara.


\section{Performance analysis}
\label{sec:performance-analysis}

In this section, we analyze Ara in terms of its peak performance
across several design parameters. We use the matrix multiplication
kernel to explore architectural limitations in depth, before analyzing
how such limitations manifest themselves for the other kernels.

\subsection{Matrix multiplication}
\label{sec:matr-mult}

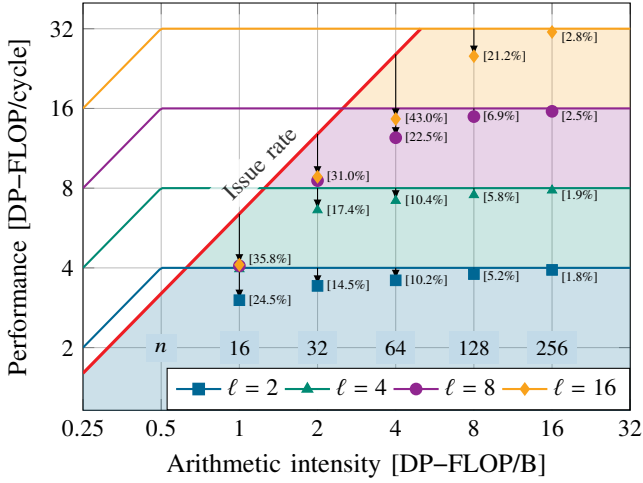
\begin{figure}[ht]
  \centering
  \begin{tikzpicture}
    \begin{axis}[
      xlabel = {Arithmetic intensity [\si{\double\flop\per\byte}]},
      log basis x = {2},
      xmode = log,
      xmin = .25,
      xmax = 32,
      ylabel={Performance [\si{\double\flop\per\cycle}]},
      log basis y = {2},
      ymode = log,
      ymax = 40,
      log ticks with fixed point,
      grid = major,
      legend style = {at={(1,0)}, anchor=south east, legend columns=-1, font=\small}]

      \addlegendimage{line legend, thick, color1, mark=square*}
      \addlegendimage{line legend, thick, color2, mark=triangle*}
      \addlegendimage{line legend, thick, color3, mark=otimes*}
      \addlegendimage{line legend, thick, color4, mark=diamond*}

      \draw[fill=color1, opacity=0, fill opacity=.2]
      (axis cs:0.25,1) -- (axis cs:0.25,1.6) -- (axis cs: 0.625,4) -- (axis cs: 32,4) -- (axis cs: 32,1) -- (axis cs: 0.5,1);

      \draw[fill=color2, opacity=0, fill opacity=.2]
      (axis cs: 0.625,4) -- (axis cs: 1.25,8) -- (axis cs: 32,8) -- (axis cs: 32,4) -- (axis cs:0.625,4);

      \draw[fill=color3, opacity=0, fill opacity=.2]
      (axis cs: 1.25,8) -- (axis cs: 2.5,16) -- (axis cs: 32,16) -- (axis cs: 32,8) -- (axis cs:1.25,8);

      \draw[fill=color4, opacity=0, fill opacity=.2]
      (axis cs: 2.5,16) -- (axis cs: 5,32) -- (axis cs: 32,32) -- (axis cs: 32,16) -- (axis cs:2.5,16);

      \addplot [colorAlert, forget plot, very thick, domain=0.25:5, name path=issue] {6.4*x};

      \addplot[color1, thick, domain=0.25:32, samples=201, name path=roof2]{roof(x,8,4)};
      \addlegendentry{$\ell = 2$}

      \addplot[color2, thick, domain=0.25:32, samples=201, name path=roof4]{roof(x,16,8)};
      \addlegendentry{$\ell = 4$}

      \addplot[color3, thick, domain=0.25:32, samples=201, name path=roof8]{roof(x,32,16)};
      \addlegendentry{$\ell = 8$}

      \addplot[color4, thick, domain=0.25:32, samples=201, name path=roof16]{roof(x,64,32)};
      \addlegendentry{$\ell = 16$}

      \node [fill=color1!17] at (axis cs: .5, 2) {\footnotesize$n$};
      \node [fill=color1!17] at (axis cs:  1, 2) {\footnotesize$16$};
      \node [fill=color1!17] at (axis cs:  2, 2) {\footnotesize$32$};
      \node [fill=color1!17] at (axis cs:  4, 2) {\footnotesize$64$};
      \node [fill=color1!17] at (axis cs:  8, 2) {\footnotesize$128$};
      \node [fill=color1!17] at (axis cs: 16, 2) {\footnotesize$256$};

      \draw [->] (axis cs: 1, 4) -- (axis cs: 1, 3.02) node [right] {\tiny[{24.5\%}]};
      \draw [->] (axis cs: 1, 6.4) -- (axis cs: 1, 4.11) node [right, pos=.9] {\tiny[{35.8\%}]};

      \draw [->] (axis cs: 2, 4) -- (axis cs: 2, 3.42) node [right] {\tiny[{14.5\%}]};
      \draw [->] (axis cs: 2, 8) -- (axis cs: 2, 6.61) node [right] {\tiny[{17.4\%}]};
      \draw [->] (axis cs: 2, 12.8) -- (axis cs: 2, 8.83) node [right] {\tiny[{31.0\%}]};

      \draw [->] (axis cs: 4, 4) -- (axis cs: 4, 3.59) node [right] {\tiny[{10.2\%}]};
      \draw [->] (axis cs: 4, 8) -- (axis cs: 4, 7.17) node [right] {\tiny[{10.4\%}]};
      \draw [->] (axis cs: 4, 16) -- (axis cs: 4, 12.4) node [right] {\tiny[{22.5\%}]};
      \draw [->] (axis cs: 4, 25.6) -- (axis cs: 4, 14.6) node [right] {\tiny[{43.0\%}]};

      \draw (axis cs: 8, 4) -- (axis cs: 8, 3.79) node [right, yshift=-.1em] {\tiny[{5.2\%}]};
      \draw (axis cs: 8, 8) -- (axis cs: 8, 7.54) node [right, yshift=-.1em] {\tiny[{5.8\%}]};
      \draw (axis cs: 8, 16) -- (axis cs: 8, 14.9) node [right] {\tiny[{6.9\%}]};
      \draw [->] (axis cs: 8, 32) -- (axis cs: 8, 25.2) node [right] {\tiny[{21.2\%}]};

      \draw (axis cs: 16, 4) -- (axis cs: 16, 3.93) node [right, yshift=-.3em] {\tiny[{1.8\%}]};
      \draw (axis cs: 16, 8) -- (axis cs: 16, 7.85) node [right, yshift=-.2em] {\tiny[{1.9\%}]};
      \draw (axis cs: 16, 16) -- (axis cs: 16, 15.6) node [right, yshift=-.2em] {\tiny[{2.5\%}]};
      \draw (axis cs: 16, 32) -- (axis cs: 16, 31.1) node [right, yshift=-.2em] {\tiny[{2.8\%}]};

      \node [rotate=45, fill=white, opacity=.8] at (axis cs: 1.2, 10) {\small Issue rate};

      \addplot [only marks, mark=square*, color1] table {matmul_2l};
      \addplot [only marks, mark=triangle*, color2] table {matmul_4l};
      \addplot [only marks, mark=otimes*, thick, color3] table {matmul_8l};
      \addplot [only marks, mark=diamond*, thick, color4] table {matmul_16l};
    \end{axis}
  \end{tikzpicture}
  \caption{Performance results for the matrix multiplication
    $C \gets AB + C$, with different number of lanes $\ell$, for
    several $n \times n$ problem sizes. The bold red line depicts a
    performance boundary due to the instruction issue rate. The
    numbers between brackets indicate the performance loss, with
    respect to the theoretically achievable peak performance.}
  \label{fig:matmul_perf}
\end{figure}

\figref{fig:matmul_perf} shows the performance measurements of the
matrix multiplication $C \gets AB + C$, for several Ara instances and
problem sizes $n \times n$. For problems ``large enough,'' the
performance results meet the peak performance boundary. For a matrix
multiplication of size \num{256 x 256}, we utilize the \glspl{FPU} for
\num{98}{\percent} of the time for an Ara instance with two lanes and
for \num{97}{\percent} for 16 lanes, \review{comparable to Hwacha's
  \num{95}$+${\percent}~\cite{Schmidt2018} and Beldianu and Ziavras's
  \num{97}{\percent}~\cite{Beldianu2014} functional units'
  utilization}. The performance scalability comes, however, at a
price. More lanes require larger problem sizes to fully exploit the
maximum performance, even though all problem sizes fall into the
compute-bound regime. Smaller problems, however, cannot fully utilize
the functional units. It is important to note that this limiting
effect can also be observed in other vector processors such as Hwacha
(see comparison in \secref{sec:hwacha-comp}).

This effect is attributed to two main reasons: first, the
initialization of the vector register file before starting
computation; and second, the rate at which the vector instructions are
issued to Ara. The former is analyzed in detail in
\apxref{sec:impl-exec-matr}. The latter is related to the rate at
which the vector \gls{FMA} instructions are issued. To understand
this, consider that smaller vectors occupy the pipeline for fewer
cycles, and more vector instructions are required to fully utilize the
\glspl{FPU}. If every vector \gls{FMA} instruction occupies the
\glspl{FPU} for $\tau$ cycles and they are issued every $\delta$
cycles, the system performance $\omega$ is limited by
\begin{equation}
  \label{eq:1}
  \omega \le \Pi \frac{\tau}{\delta}.
\end{equation}
For the $n \times n$ matrix multiplication, $\tau$ is equal to
$2n/\Pi$. We use this together with Equation~(\ref{eq:3}) to rewrite
this constraint in terms of the arithmetic intensity
$I_{\text{MATMUL}}$, resulting in
\begin{equation}
  \label{eq:2}
  \omega \le \frac{32}{\delta} I_{\text{MATMUL}}.
\end{equation}
This translates to another performance boundary in the roofline plot,
purely dependent on the instruction issue rate. The \gls{FMA}
instructions are issued every five cycles, as discussed in
\apxref{sec:impl-exec-matr}. This shifts the roofline of the
architecture as illustrated with the bold line in
\figref{fig:matmul_perf}. Note that, for 16 lanes, even the
performance of a \num{64 x 64} matrix multiplication ends up being
limited by the vector instruction issue rate.

The performance degradation with shorter vectors could be mitigated
with a more complex instruction issue mechanism, either going
superscalar or introducing a \gls{VLIW} capable \gls{ISA} to increase
the issue rate. Shorter vectors bring vector processors to an array
processor, where the vector instructions execute for a single
cycle. This puts pressure on the issue logic, demanding more than a
simple single-issue in-order core. For example, all ARM Cortex-A cores
with Neon capability are also superscalar~\cite{ArmCortex2019}.
Another alternative would be the use of a \gls{MIMD} approach where
the lanes would be decoupled, running instructions issued by different
scalar cores, \review{as discussed by Lu~\etal\cite{Lu2016}. While
  fine-grain temporal sharing of the vector units achieves an exciting
  increase of the \gls{FPU} utilization~\cite{Lu2016},} duplication of
the instruction issue logic could also degrade the energy efficiency
achieved by the design.

\subsection{AXPY}
\label{sec:daxpy}

As discussed in \secref{sec:benchmarks}, DAXPY is a heavily
memory-bound kernel, with an arithmetic intensity of
\SI{0.083}{\double\flop\per\byte}. It is no surprise that the measured
performance for such a kernel are much less than the system's peak
performance in the compute-bound region. For an Ara instance with two
lanes, we measure \SI{0.65}{\double\flop\per\cycle}, which is
\num{98}{\percent} of the theoretical performance limit. For sixteen
lanes, the achieved \SI{4.27}{\double\flop\per\cycle} is still
\num{80}{\percent} of the theoretical limit $\beta I_{\text{DAXPY}}$
from the roofline plot. The limiting factor is the configuration of
the vector unit, whose overhead increases the runtime from the ideal
\num{96} cycles to \num{120} cycles.

\subsection{Convolution}
\label{sec:convolution}

Convolutions are heavily compute-bound kernels, with an arithmetic
intensity up to of \SI{34.9}{\double\flop\per\byte}. With two lanes,
it achieves a performance up to \SI{3.73}{\double\flop\per\cycle}. We
notice some performance degradation for sixteen lanes, where the
kernel achieves \SI{26.7}{\double\flop\per\cycle}, \ie an \gls{FPU}
utilization of \num{83.2}{\percent}, close to the performance achieved
by the \num{128x128} matrix multiplication. The reason for the
performance drop at both kernels lies in the problem size. In this
case, each lane holds only seven elements of the \num{112}-element
long vectors, \ie the vectors do not even occupy the eight banks. With
such short instructions, the system does not have enough time to
achieve the steady state banking access pattern discussed in
\secref{sec:vector-register-file}. Such short instructions also incur
into banking conflicts that would otherwise be amortized across longer
vectors.

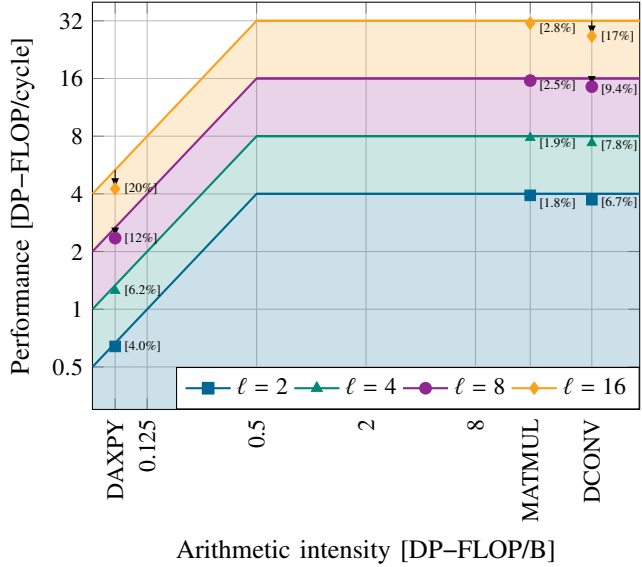
\begin{figure}[ht]
  \centering
  \begin{tikzpicture}
    \begin{axis}[
      xlabel = {Arithmetic intensity [\si{\double\flop\per\byte}]},
      log basis x = {2},
      xmode = log,
      xmin = .0625,
      xmax = 64,
      xtick = \empty,
      xticklabel style = {rotate=90},
      extra x ticks = {0.08333, 16, 34.9, 0.5, 0.125, 2, 8},
      extra x tick labels = {DAXPY, MATMUL, DCONV, 0.5, 0.125, 2, 8},
      ylabel={Performance [\si{\double\flop\per\cycle}]},
      log basis y = {2},
      ymode = log,
      ymin = .3,
      ymax = 40,
      log ticks with fixed point,
      grid = major,
      legend style = {at={(1,0)}, anchor=south east, legend columns=-1, font=\small}]

      \addlegendimage{line legend, thick, color1, mark=square*}
      \addlegendimage{line legend, thick, color2, mark=triangle*}
      \addlegendimage{line legend, thick, color3, mark=otimes*}
      \addlegendimage{line legend, thick, color4, mark=diamond*}

      \addplot[color1, thick, domain=0.0625:64, samples=201, name path=roof2]{roof(x,8,4)};
      \addplot[color1, thick, domain=0.500:64, name path=compA] {4};
      \addlegendentry{$\ell = 2$}
      \path[name path=axisAa] (axis cs:0.0625,.3) -- (axis cs:0.5,4);
      \path[name path=axisBa] (axis cs:0.0625,.3) -- (axis cs:0.0625,0.5);
      \path[name path=axisCa] (axis cs:0.0625,.3) -- (axis cs:64,.3);
      \addplot[fill=color1, fill opacity=.2] fill between [of=axisAa and axisBa];
      \addplot[fill=color1, fill opacity=.2] fill between [of=compA  and axisCa];

      \addplot[color2, thick, domain=0.0625:64, samples=201, name path=roof4]{roof(x,16,8)};
      \addplot[color2, thick, domain=0.500:64, name path=compB] {8};
      \addlegendentry{$\ell = 4$}
      \path[name path=axisAb] (axis cs:0.0625,.5) -- (axis cs:0.0625,1);
      \path[name path=axisBb] (axis cs:0.5,4) -- (axis cs:0.5,8);
      \addplot[fill=color2, fill opacity=.2] fill between [of=axisAb and axisBb];
      \addplot[fill=color2, fill opacity=.2] fill between [of=compB and compA];

      \addplot[color3, thick, domain=0.0625:64, samples=201, name path=roof8]{roof(x,32,16)};
      \addplot[color3, thick, domain=0.500:64, name path=compC] {16};
      \addlegendentry{$\ell = 8$}
      \path[name path=axisAc] (axis cs:0.0625,1) -- (axis cs:0.0625,2);
      \path[name path=axisBc] (axis cs:0.5,8) -- (axis cs:0.5,16);
      \addplot[fill=color3, fill opacity=.2] fill between [of=axisAc and axisBc];
      \addplot[fill=color3, fill opacity=.2] fill between [of=compC and compB];

      \addplot[color4, thick, domain=0.0625:64, samples=201, name path=roof16]{roof(x,64,32)};
      \addplot[color4, thick, domain=0.500:64, name path=compD] {32};
      \addlegendentry{$\ell = 16$}
      \path[name path=axisAd] (axis cs:0.0625,2) -- (axis cs:0.0625,4);
      \path[name path=axisBd] (axis cs:0.5,16) -- (axis cs:0.5,32);
      \addplot[fill=color4, fill opacity=.2] fill between [of=axisAd and axisBd];
      \addplot[fill=color4, fill opacity=.2] fill between [of=compD and compC];

      \draw (axis cs: 0.083333, 0.66667) -- (axis cs: 0.083333, 0.64) node [right] {\tiny[{4.0\%}]};
      \draw (axis cs: 0.083333, 1.33333) -- (axis cs: 0.083333, 1.25) node [right] {\tiny[{6.2\%}]};
      \draw [->] (axis cs: 0.083333, 2.66667) -- (axis cs: 0.083333, 2.35) node [right] {\tiny[{12\%}]};
      \draw [->] (axis cs: 0.083333, 5.33333) -- (axis cs: 0.083333, 4.25) node [right] {\tiny[{20\%}]};

      \draw (axis cs: 16, 4) -- (axis cs: 16, 3.93) node [right, yshift=-.3em] {\tiny[{1.8\%}]};
      \draw (axis cs: 16, 8) -- (axis cs: 16, 7.85) node [right, yshift=-.2em] {\tiny[{1.9\%}]};
      \draw (axis cs: 16, 16) -- (axis cs: 16, 15.6) node [right, yshift=-.2em] {\tiny[{2.5\%}]};
      \draw (axis cs: 16, 32) -- (axis cs: 16, 31.1) node [right, yshift=-.2em] {\tiny[{2.8\%}]};

      \draw (axis cs: 34.9, 4) -- (axis cs: 34.9, 3.73) node [right, yshift=-.1em] {\tiny[{6.7\%}]};
      \draw (axis cs: 34.9, 8) -- (axis cs: 34.9, 7.37) node [right, yshift=-.1em] {\tiny[{7.8\%}]};
      \draw [->] (axis cs: 34.9, 16) -- (axis cs: 34.9, 14.5) node [right, yshift=-.1em] {\tiny[{9.4\%}]};
      \draw [->] (axis cs: 34.9, 32) -- (axis cs: 34.9, 26.6) node [right] {\tiny[{17\%}]};

      \addplot [only marks, mark=square*, color1] coordinates {(0.083333, 0.64) (16,3.93) (34.9, 3.73)};
      \addplot [only marks, mark=triangle*, color2] coordinates {(0.083333, 1.25) (16,7.85) (34.9, 7.37)};
      \addplot [only marks, mark=otimes*, thick, color3] coordinates {(0.083333, 2.35) (16,15.6) (34.9, 14.5)};
      \addplot [only marks, mark=diamond*, thick, color4] coordinates {(0.083333, 4.25) (16,31.1) (34.9, 26.6)};
    \end{axis}
  \end{tikzpicture}
  \caption{Performance results for the three considered benchmarks,
    with different number of lanes $\ell$. AXPY uses vectors of length
    \num{256}, the MATMUL is between matrices of size \num{256x256},
    and CONV uses GoogLeNet's sizes. The numbers between brackets
    indicate the performance loss, with respect to the theoretically
    achievable peak performance.}
  \label{fig:kern_perf}
\end{figure}

\figref{fig:kern_perf} shows the performance results for the three
considered benchmarks. In both memory- and compute-bound regions, the
achieved performance tends to achieve the roofline boundary, for all
the considered architecture instances.

\subsection{Performance comparison with Hwacha}
\label{sec:hwacha-comp}

For comparison with Ara, we measured Hwacha's performance for the
matrix multiplication benchmark, using the publicly available
\gls{HDL} sources and tooling scripts from their GitHub
repository\footnote{See
  https://github.com/ucb-bar/hwacha-template/tree/a5ed14a.}. We were
not able to reproduce the \num{32 x 32} double precision matrix
multiplication performance claimed by
Dabbelt~\etal\cite{Dabbelt2016}. This is because Hwacha relies on a
closed-source L2 cache, whereas its public version has a limited
memory system with no banked cache and a broadcast hub to ensure
coherence. This effectively limits Hwacha's memory bandwidth to
\SI{128}{\bit\per\cycle}, starving the \gls{FMA} units and capping the
achievable performance.

\tabref{tab:comparison_ara_hwacha} brings the performance achieved by
Ara and the published results for Hwacha~\cite{Dabbelt2016} side by
side. For a fair comparison, the roofline boundaries are identical
between the compared architectures. For small problems, for which a
direct comparison is possible, Ara utilizes its \glspl{FPU} much
better than the equivalent Hwacha instances. For the instances with
two lanes, Ara utilizes its \glspl{FPU} \num{66}{\percent} more than
the equivalent Hwacha instance, for a relatively small \num{32x32}
matrix multiplication. Moreover, we note that both Ara and Hwacha
operate at a similar architectural design point in the sense that they
are coupled to a single-issue in-order core. Therefore, Hwacha
exhibits a similar performance degradation on small matrices and
vector lengths as previously described for Ara in
\secref{sec:matr-mult}. For what concerns large problems, another more
recent reference on Hwacha~\cite{Schmidt2018} claims a
\num{95}{\percent} \gls{FPU} utilization for a \num{128 x 128} MATMUL,
close to the performance level that Ara achieves. However, these
results cannot be reproduced on the current open-source version of
Hwacha, possibly due to the memory system limitation outlined above.

\begin{table}[ht]
  \caption{Normalized achieved performance between equivalent Ara
    and Hwacha instances for a matrix multiplication, with different
    $n \times n$ problem sizes.}
  \label{tab:comparison_ara_hwacha}
  {\centering
  \setlength{\tabcolsep}{5pt}
  \begin{tabular}{rcccccc}
    \toprule
    $\Pi$ & \multicolumn{2}{c}{\SI{8}{\double\flop\per\cycle}} & \multicolumn{2}{c}{\SI{16}{\double\flop\per\cycle}} & \multicolumn{2}{c}{\SI{32}{\double\flop\per\cycle}} \\\cmidrule(lr){1-1} \cmidrule(lr){2-3} \cmidrule(lr){4-5} \cmidrule(lr){6-7}
    $n$   & Ara                  & Hwacha$^a$                  & Ara                  & Hwacha               & Ara                  & Hwacha               \\\midrule
    16    & \num{49.5}{\percent} & ---                         & \num{25.4}{\percent} & ---                  & \num{12.8}{\percent} & ---                  \\
    32    & \num{82.6}{\percent} & \num{49.9}{\percent}        & \num{53.4}{\percent} & \num{35.6}{\percent} & \num{27.6}{\percent} & \num{22.4}{\percent} \\
    64    & \num{89.6}{\percent} & ---                         & \num{77.5}{\percent} & ---                  & \num{45.6}{\percent} & ---                  \\
    128   & \num{94.3}{\percent} & ---                         & \num{93.1}{\percent} & ---                  & \num{78.8}{\percent} & ---                  \\\bottomrule
  \end{tabular}}

  \vspace{.5em}
  \footnotesize $^a$Performance results extracted
  from~\cite{Dabbelt2016}.
\end{table}


\section{Implementation results}
\label{sec:results}

In this section, we analyze the implementation of several Ara
instances, in terms of area, power and energy efficiency.

\subsection{Methodology}
\label{sec:phys-char}

Ara was synthesized for \textsc{GlobalFoundries} 22FDX \gls{FDSOI}
technology using Synopsys Design Compiler 2017.09. The back-end design
flow was carried out with Cadence Innovus 18.11.000. For this
technology, one gate equivalent (\si{\gate}) is equal to
\SI{0.199}{\micro\meter\squared}. Ara's performance and power figures
of merit are measured running the kernels on a cycle-accurate
\gls{RTL} simulation. We used Synopsys PrimeTime 2016.12 to extract
the power figures \review{with activities obtained with timing
  information from the implemented design at
  TT/\SI{0.80}{\volt}\kern-0.12em/\SI{25}{\celsius}}. \tabref{tab:parameters}
summarizes Ara's design parameters.

\begin{table}[ht]
  \centering
  \caption{Design parameters.}
  \label{tab:parameters}
  \begin{tabular}[ht]{rrl}
    \toprule
                                                          & \#~Lanes            & $\ell \in [2, 4, 8, 16]$                          \\
                                                          & Memory width        & $32\ell~\si{\bit}$                                \\
                                                          & Operating corner    & TT/\SI{0.80}{\volt}\kern-0.12em/\SI{25}{\celsius} \\
                                                          & Target frequency    & \SI{1}{\giga\hertz}                               \\\midrule
    \multirow{3}{*}{\rotatebox[origin=c]{90}{\emph{VRF}}} & Size                & \SI{16}{\kibi\byte\per\lane}                      \\
                                                          & \#~Banks            & \SI{8}{\bank\per\lane}                            \\
                                                          & Bank width          & \SI{64}{\bit}                                     \\\bottomrule
  \end{tabular}
\end{table}

Because the maximum frequencies achieved after synthesis are usually
higher than the ones achieved after the back-end flow, the system was
synthesized for a clock period constraint \SI{250}{\pico\second}
shorter than the target clock period of \SI{1}{\nano\second}. The
system can be tuned for even higher frequencies by deploying \gls{FBB}
techniques, at the expense of an increase in leakage power. In
average, the final designs have a mix of \num{72.9}{\percent}
\gls{LVT} cells and \num{27.1}{\percent} \gls{SLVT} cells.

\subsection{Physical implementation}
\label{sec:phys-impl}

\begin{figure}[ht]
  \centering
  \begin{minipage}[ht]{\linewidth}
    \centering \subfloat[Place-and-route results of an Ara instance
    with four lanes, highlighting its internal blocks: A) lane 0; B)
    lane 1; C) lane 2; D) lane 3; E) \gls{SLDU}; F) sequencer; G)
    \gls{VLSU}; H) Ara front end; I) Ariane; J) memory interconnect.]{
      \begin{tikzpicture}
        \node[anchor=south west,inner sep=0] (image) at (0,0) {\includegraphics[width=.99\linewidth]{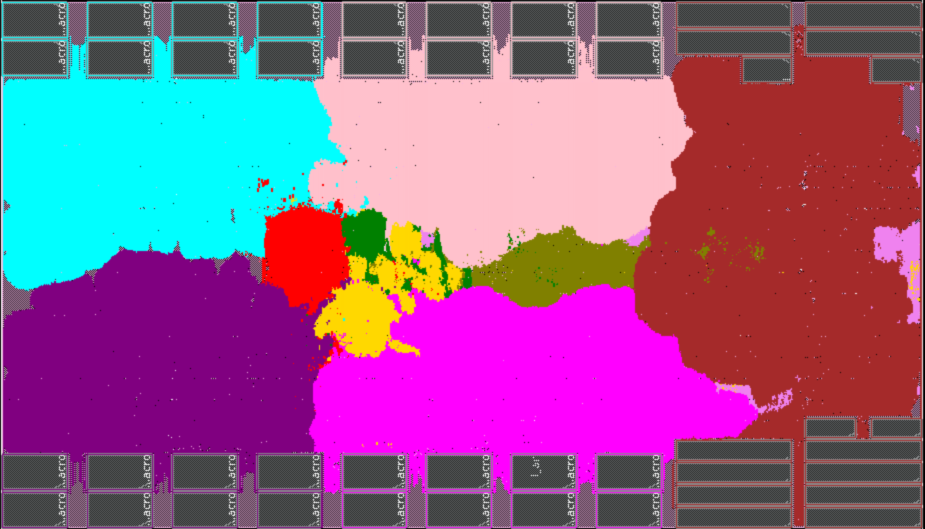}};
        \begin{scope}[x={(image.south east)}, y={(image.north west)}]
          \node [thick] at (0.166,0.7039) {\textbf{A}};
          \node [thick,white] at (0.166,0.2868) {\textbf{B}};
          \node [thick] at (0.526,0.6981) {\textbf{C}};
          \node [thick, white] at (0.566,0.2788) {\textbf{D}};
          \node [thick, white] at (0.330,0.5237) {\textbf{E}};
          \node [thick, white] at (0.394,0.5538) {\textbf{F}};
          \node [thick] at (0.39,0.41) {\textbf{G}};
          \node [thick, white] at (0.595,0.5) {\textbf{H}};
          \node [thick, white] at (0.834,0.5066) {\textbf{I}};
          \node [thick, white] at (0.978,0.5027) {\textbf{J}};
        \end{scope}
      \end{tikzpicture}}
  \end{minipage}

  \vspace{1em}

  \begin{minipage}[ht]{\linewidth}
    \centering \subfloat[Detail of one of Ara's lanes, highlighting
    its internal blocks: A) lane sequencer; B) \gls{VRF}; C) operand
    queues; D) \gls{MUL}; E) \gls{FPU}; F) \gls{ALU}.]{
      \begin{tikzpicture}
        \node[anchor=south west,inner sep=0] (image) at (0,0) {\includegraphics[width=.99\linewidth]{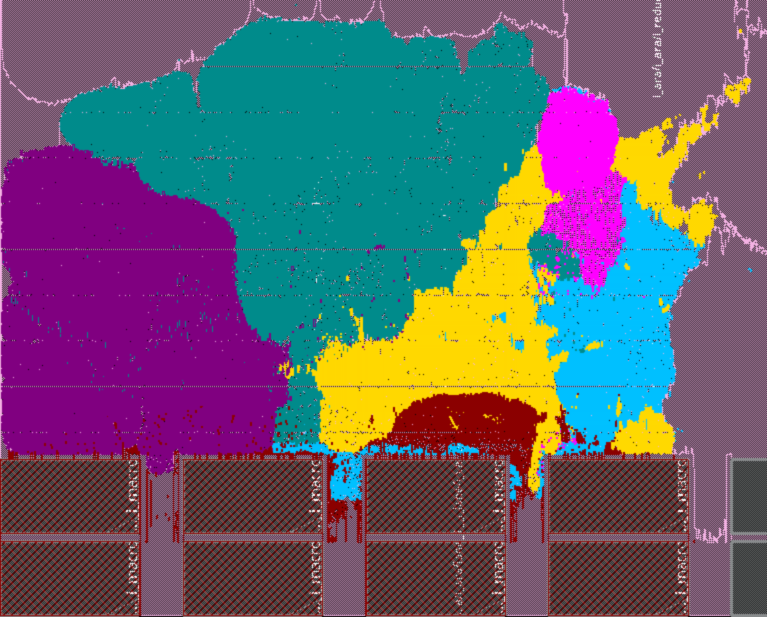}};
        \begin{scope}[x={(image.south east)}, y={(image.north west)}]
          \node [thick] at (0.8,0.4686) {\textbf{A}};
          \node [thick, white] at (0.45,0.13) {\textbf{B}};
          \node [thick] at (0.6,0.4582) {\textbf{C}};
          \node [thick, white] at (0.18,0.45) {\textbf{D}};
          \node [thick, white] at (0.45,0.7) {\textbf{E}};
          \node [thick, white] at (0.764,0.6986) {\textbf{F}};
        \end{scope}
      \end{tikzpicture}}
  \end{minipage}

  \caption{Place-and-route results of an Ara instance with four lanes
    in \textsc{GlobalFoundries} \SI{22}{\nano\meter} technology on a
    $\SI{1.125}{\mm} \times \SI{1.000}{\mm}$ macro.}
  \label{fig:impl}
\end{figure}
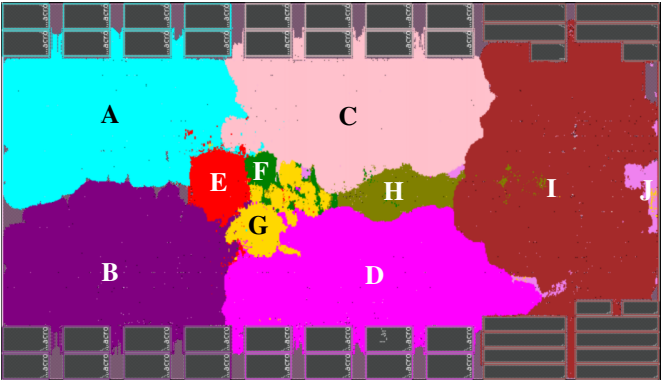
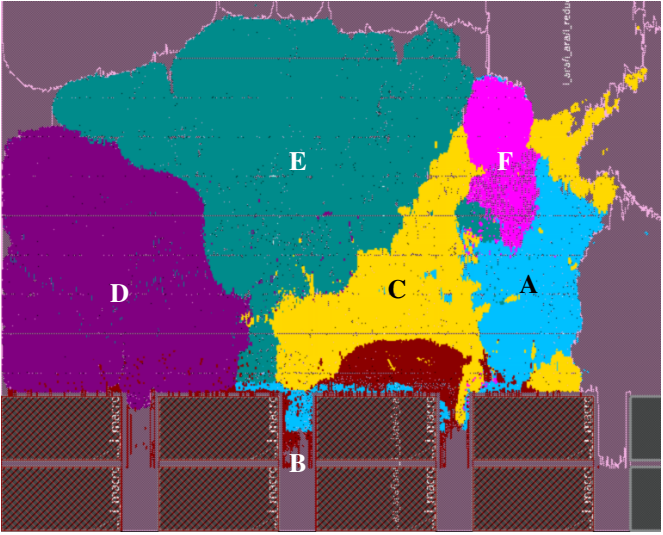

We implemented four Ara instances, with two, four, eight and sixteen
lanes. The instance with four lanes was placed and routed as a
$\SI{1.125}{\mm} \times \SI{1.000}{\mm}$ macro in
\textsc{GlobalFoundries} 22FDX \gls{FDSOI} technology, using Cadence
Innovus~18.11.000. \figref{fig:impl} shows the final implemented
result, highlighting its internal blocks. Without its caches, Ariane
uses about the same area (\SI{524}{\kilo\gate}) as lane, including its
\gls{VRF}.

Our vector processor is scalable, in the sense that Ariane can be
reused without changes to drive a wide range of different lane
parameters. Furthermore, each vector lane touches only its own section
of the \gls{VRF}, hence it does not introduce any scalability
bottlenecks. Scalability is only limited by the units that need to
interface with all lanes at once, namely the main sequencer, the
\gls{VLSU}, and the \gls{SLDU}. Beldianu and
Ziavras~\cite{Beldianu2014} and Hwacha~\cite{Schmidt2018}, on the
other hand, have a dedicated memory port per lane. This solves the
scalability issue locally, by controlling the growth of the memory
interface, but pushes the memory interconnect issue further upstream,
as its wide memory system must be able to aggregate multiple parallel
requests from all these ports to achieve their maximum memory
throughput.

\review{We decided not to deploy lane-level \gls{PG} or \gls{BB}
  techniques, due to their significant area and timing impact. In
  terms of area, both techniques would require an isolation ring
  \SI{10}{\um}-wide around each \gls{PG}/\gls{BB} domain, or at least
  an \num{8}{\percent} increase in the area of each lane. In terms of
  timing, isolation cells between power domains and separated clock
  trees would impact Ara's operating frequency. Assuming these cells
  would be in the critical path between the lanes and the \gls{VLSU},
  this would incur into a \num{10}{\percent} clock frequency
  penalty. Reverse Body-Biasing lowers the leakage, but also impacts
  frequency, since it cannot be applied to high-performance \gls{LVT}
  and \gls{SLVT} cells. Furthermore, \gls{PG} (and, to a lesser
  degree, \gls{BB}) would introduce significant (in the order of
  $\num{10}-\num{15}$ cycles) turn-on transition times, which could be
  tolerable only if coupled with a scheduling policy for power
  managing the lanes. These techniques are out of the scope of the
  current work.}

\subsection{Performance, power, and area results}
\label{sec:perf-power-area}

\tabref{tab:energy} summarizes the post-place-and-route results of
several Ara instances. Overall, the instances achieve nominal
operating frequencies around \SI{1.2}{\giga\hertz}, \review{where we
  chose the typical corner,
  TT/\SI{0.80}{\volt}\kern-0.12em/\SI{25}{\celsius}, for comparison
  with equivalent results from Hwacha~\cite{Lee2015b}. For
  completeness, \tabref{tab:energy} also presents timing results for
  the worst-case corner, \ie
  SS/\SI{0.72}{\volt}\kern-0.11em/\SI{125}{\celsius}.}

\begin{table*}[t]
  \caption{Post-place-and-route architectural comparison between
    several Ara instances in \textsc{GlobalFoundries} 22FDX
    \gls{FDSOI} technology in terms of performance, power consumption,
    and energy efficiency.}
  \label{tab:energy}

  \aboverulesep=-.05ex
  \belowrulesep=-.05ex
  \renewcommand{\arraystretch}{1.8}
  \setlength{\tabcolsep}{5.35pt}

  {\centering
   \begin{tabular}{@{}l|lll|lll|lll|lll@{}}
     \toprule
                                                    & \multicolumn{12}{l}{\textbf{Instance}}                                                                                                \\
     \textbf{Figure of merit}                       & \multicolumn{3}{l}{$\ell = 2$}  & \multicolumn{3}{l}{$\ell = 4$}  & \multicolumn{3}{l}{$\ell = 8$}  & \multicolumn{3}{l}{$\ell = 16$} \\\cmidrule{1-13}
     \textbf{Clock} (nominal) [\si{\giga\hertz}]    & \multicolumn{3}{l|}{\num{1.25}} & \multicolumn{3}{l|}{\num{1.25}} & \multicolumn{3}{l|}{\num{1.17}} & \multicolumn{3}{l}{\num{1.04}}  \\\cmidrule{1-13}
     \textbf{Clock} (worst-case) [\si{\giga\hertz}] & \multicolumn{3}{l|}{\num{0.92}} & \multicolumn{3}{l|}{\num{0.93}} & \multicolumn{3}{l|}{\num{0.87}} & \multicolumn{3}{l}{\num{0.78}}  \\\cmidrule{1-13}
     \textbf{Area} [\si{\kilo\gate}]                & \multicolumn{3}{l|}{\num{2228}} & \multicolumn{3}{l|}{\num{3434}} & \multicolumn{3}{l|}{\num{5902}} & \multicolumn{3}{l}{\num{10735}} \\
     \emph{Area per lane} [\si{\kilo\gate}]         & \multicolumn{3}{l|}{\num{1114}} & \multicolumn{3}{l|}{\num{858}}  & \multicolumn{3}{l|}{\num{738}}  & \multicolumn{3}{l}{\num{671}}   \\\toprule

     \textbf{Kernel}                                & \emph{matmul}$^a$  & \emph{dconv}$^b$   & \emph{daxpy}$^c$  & \emph{matmul}      & \emph{dconv}       & \emph{daxpy}      & \emph{matmul}      & \emph{dconv}       & \emph{daxpy}       & \emph{matmul}      & \emph{dconv}       & \emph{daxpy}       \\\cmidrule{1-13}
     \textbf{Performance} [\si{\dpgiga\flops}]      & \num{4.91}         & \num{4.66}         & \num{0.82}        & \num{9.80}         & \num{9.22}         & \num{1.56}        & \num{18.2}         & \num{16.9}         & \num{2.80}         & \num{32.4}         & \num{27.7}         & \num{4.44}         \\\cmidrule{1-13}
     \textbf{Core power} [\si{\milli\watt}]         & \num{138}          & \num{130}          & \num{68.2}        & \num{259}          & \num{239}          & \num{113}         & \num{456}          & \num{420}          & \num{183}          & \num{794}          & \num{676}          & \num{280}          \\
     Leakage [\si{\milli\watt}]                     & \num{7.2}          &                    &                   & \num{11.2}         &                    &                   & \num{21.1}         &                    &                    & \num{31.4}         &                    &                    \\
     Ariane/Ara [\si{\milli\watt}]                  & \num{22}/\num{116} & \num{22}/\num{108} & \num{20}/\num{48} & \num{27}/\num{232} & \num{29}/\num{210} & \num{25}/\num{88} & \num{28}/\num{428} & \num{29}/\num{391} & \num{24}/\num{159} & \num{31}/\num{763} & \num{31}/\num{646} & \num{25}/\num{255} \\
     \emph{Core power per lane} [\si{\milli\watt}]  & \num{69}           & \num{65}           & \num{34}          & \num{65}           & \num{60}           & \num{28}          & \num{57}           & \num{54}           & \num{23}           & \num{50}           & \num{42}           & \num{15}           \\\cmidrule{1-13}
     \textbf{Efficiency} [\si{\dpgiga\flops\per\W}] & \num{35.6}         & \num{35.8}         & \num{12.0}        & \num{37.8}         & \num{38.6}         & \num{13.8}        & \num{39.9}         & \num{40.2}         & \num{15.3}         & \num{40.8}         & \num{41.0}         & \num{15.9}         \\\bottomrule
   \end{tabular}}

 \vspace{1ex}
 \footnotesize$^a$Double precision floating point \num{256 x 256}
  matrix multiplication.  $^b$Double precision floating point tensor
  convolution with sizes from the first layer of GoogLeNet. Input size
  is \num{3 x 112 x 112} and kernel size is \num{64 x 3 x 7 x
    7}. $^c$Double precision AXPY of vectors with length
  \num{256}.
\end{table*}

The two-lane instance has its critical path inside the double
precision \gls{FMA}. This block relies on the automatic retiming
feature from Synopsys Design Compiler, and the register placement
could be further improved by hand-tuning, or by increasing the number
of pipeline stages. Another critical path is on the combinational
handshake between the \gls{VLSU} and its operand queues in the
lanes. Both paths are about \num{40} gate delays long. Timing of the
instances with eight and sixteen lanes becomes increasingly critical,
due to the widening of Ara's memory interface. This happens when the
\gls{VLSU} collects \SI{64}{\bit} words from all the lanes, realigns
and packs them into a wide word to be sent to memory. The instance
with \num{16} lanes incurs into a \num{17}{\percent} clock frequency
penalty when compared with the frequency achieved by the instance with
two lanes.

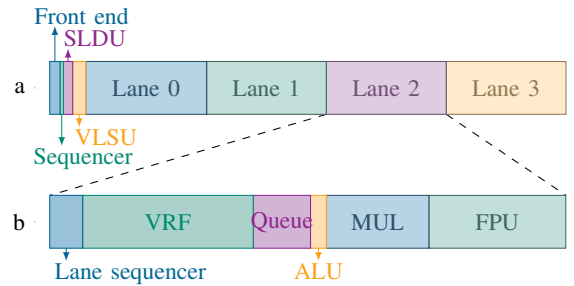
\begin{figure}[ht]
  \centering
  \begin{tikzpicture}[/tikz/font=\small]
    \begin{axis}[
      xbar stacked,
      bar width = 2em,
      ytick distance = 1,
      y axis line style = {draw=none},
      axis x line = none,
      tickwidth = 0pt,
      enlarge x limits = .03,
      enlarge y limits = .6,
      height = 5.5cm,
      nodes near coords,
      symbolic y coords = {b,a},
      ytick={b,a},
      every pin/.style={font=\footnotesize},
      point meta = explicit symbolic]

      \def\fend{2.00}
      \def\seq{0.65}
      \def\sldu{1.79}
      \def\vlsu{2.52}
      \def\laneA{23.39}
      \def\laneB{23.13}
      \def\laneC{23.21}
      \def\laneD{23.16}

      \def\lseq{6.37}
      \def\vrf{32.99}
      \def\opq{11.08}
      \def\alu{ 3.08}
      \def\mul{19.81}
      \def\fpu{26.48}


      \addplot+[xbar] plot coordinates {(0,a) (0,b)};
      \addplot+[xbar, color=color1, fill=color1!30] plot coordinates {(\fend,a) (\lseq,b)};
      \addplot+[xbar, color=color2, fill=color2!30] plot coordinates {(\seq,a) (\vrf,b) [\gls{VRF}]};
      \addplot+[xbar, color=color3, fill=color3!30] plot coordinates {(\sldu,a) (\opq,b) [Queue]};
      \addplot+[xbar, color=color4, fill=color4!30] plot coordinates {(\vlsu,a) (\alu,b)};
      \addplot+[xbar, color=color1!50!black, fill=color1!20] plot coordinates {(\laneA,a) [Lane 0] (\mul,b) [\gls{MUL}]};
      \addplot+[xbar, color=color2!50!black, fill=color2!20] plot coordinates {(\laneB,a) [Lane 1] (\fpu,b) [\gls{FPU}]};
      \addplot+[xbar, color=color3!50!black, fill=color3!20] plot coordinates {(\laneC,a) [Lane 2]};
      \addplot+[xbar, color=color4!50!black, fill=color4!20] plot coordinates {(\laneD,a) [Lane 3]};

      \pgfmathsetmacro{\xA}{\fend/2};
      \node [pin={[pin distance=.7cm, pin edge={color1, ->}, inner sep = .5pt, color1] 90:{\hspace{5ex}Front end}}] at (axis cs:\xA,a) {};
      \pgfmathsetmacro{\xB}{\fend+\seq/2};
      \node [pin={[pin distance=.65cm, pin edge={color2, ->}, inner sep = .5pt, color2]-90:{\hspace{4ex}Sequencer}}] at (axis cs:\xB,a) {};
      \pgfmathsetmacro{\xC}{\fend+\seq+\sldu/2};
      \node [pin={[pin distance=.4cm, pin edge={color3, ->}, inner sep = .5pt, color3] 90:{\hspace{5ex}\gls{SLDU}}}] at (axis cs:\xC,a) {};
      \pgfmathsetmacro{\xD}{\fend+\seq+\sldu+\vlsu/2};
      \node [pin={[pin distance=.4cm, pin edge={color4, ->}, inner sep = .5pt, color4]-90:{\hspace{5ex}\gls{VLSU}}}] at (axis cs:\xD,a) {};
      \pgfmathsetmacro{\xE}{\lseq/2};
      \node [pin={[pin distance=.4cm, pin edge={color1, ->}, inner sep = .5pt, color1]-90:{\hspace{12ex}Lane sequencer}}] at (axis cs:\xE,b) {};
      \pgfmathsetmacro{\xF}{\lseq+\vrf+\opq+\alu/2};
      \node [pin={[pin distance=.4cm, pin edge={color4, ->}, inner sep = .5pt, color4]-90:{\gls{ALU}}}] at (axis cs:\xF,b) {};

      \pgfmathsetmacro{\xG}{\fend+\seq+\sldu+\vlsu+\laneA+\laneB};
      \pgfmathsetmacro{\xH}{\fend+\seq+\sldu+\vlsu+\laneA+\laneB+\laneC};
      \pgfmathsetmacro{\xI}{\fend+\seq+\sldu+\vlsu+\laneA+\laneB+\laneC+\laneD};
      \draw [thin, dashed] ([yshift=-1em]axis cs:\xG,a) -- ([yshift=1em]axis cs:0,b);
      \draw [thin, dashed] ([yshift=-1em]axis cs:\xH,a) -- ([yshift=1em]axis cs:\xI,b);
    \end{axis}
  \end{tikzpicture}

  \caption{Area breakdowns of a) an Ara instance with four lanes with
    detail on b) one of its lanes. Ara's total area, excluding the
    scalar processor, is \SI{2.46}{\mega\gate}. Each lane has about
    \SI{575}{\kilo\gate}.}
  \label{fig:area_breakdown}
\end{figure}

The silicon area and leakage power of the accompanying scalar core are
amortized among the lanes, which can be seen with the decreasingly
area per lane figure of merit. \figref{fig:area_breakdown} shows the
area breakdown of an Ara instance with four lanes. Ara's total area
(excluding the scalar core) is \SI{2.46}{\mega\gate}, out of which
each lane amounts to \SI{575}{\kilo\gate}. The area of the vector unit
is dominated by the lanes, while the other blocks amount to only
\num{7}{\percent} of the total area. The area of the lanes is
dominated by the \gls{VRF} (\num{35}{\percent}), the \gls{FPU}
(\num{27}{\percent}), and the multiplier (\num{18}{\percent}).

In terms of post-synthesis logic area, a Hwacha instance with four
lanes uses \SI{0.354}{\milli\meter\squared}~\cite{Dabbelt2016}, or
\SI{1098}{\kilo\gate}\footnote{As Dabbelt \etal\cite{Dabbelt2016} do
  not specify the technology they used, we considered an ideal scaling
  from \SI{28}{\nano\meter} to \SI{22}{\nano\meter}. Therefore, we
  considered one \si{\gate} in \SI{28}{\nm} to be $(28/22)^2$ bigger
  than one \si{\gate} in \SI{22}{\nm}, or
  \SI{0.322}{\micro\meter\squared}.}. When comparing post-synthesis
results, Hwacha is \num{9}{\percent} smaller than the equivalent Ara
instance. The trend is also valid for equivalent instances with eight
and sixteen lanes. The main reason for this area difference is that
Hwacha has only half as many multipliers as Ara, \ie Hwacha has one
\gls{MUL} per two \gls{FMA} units~\cite{Lee2015}. These multipliers
make up for a \num{9}{\percent} area difference. Moreover, these
Hwacha instances do not support mixed-precision
arithmetic~\cite{Dabbelt2016}, and its support would incur into a
\num{4}{\percent} area overhead~\cite{Lee2015b}. \review{Ara, however,
  has a simpler execution mechanism than Hwacha's Vector Runahead
  Unit~\cite{Lee2015}, contributing to the area difference.}

We used the placed-and-routed designs to analyze the performance and
energy efficiency of Ara when running the considered benchmarks. Due
to the asymmetry between the code that runs in Ariane and in Ara, we
extracted switching activities by running the benchmarks with netlists
back annotated with timing information. As expected, the energy
efficiency of Ara coupled to an Ariane core is considerably higher
than that of an Ariane core alone. For instance, a \num{256x256}
integer matrix multiplication achieves up to
\SI{43.6}{\giga\ops\per\watt} energy efficiency on an Ara with four
lanes, whereas a comparable benchmark runs at
\SI{17}{\giga\ops\per\watt} on Ariane~\cite{Zaruba2019}. In that case,
the instruction and data caches alone are responsible for
\num{46}{\percent} of Ariane's power dissipation. In Ara's case, most
of the memory accesses go directly into the \gls{VRF} and energy spent
for cache accesses can be amortized over many vector lanes and cycles,
increasing the system's energy efficiency with an increasing number of
lanes.

A Hwacha implementation in ST \SI{28}{\nm} \gls{FDSOI} technology (at
an undisclosed condition) achieves a peak energy efficiency of
\SI{40}{\dpgiga\flops\per\watt}~\cite{Schmidt2018}. Adjusting for
scaling gains~\cite{Dreslinski2010}, an energy efficiency of
\SI{41}{\dpgiga\flops\per\watt} is comparable to the energy efficiency
of the large Ara instances running MATMUL.


\section{Conclusions}
\label{sec:conclusions}

In this work, we presented Ara, a parametric in-order high-performance
energy-efficient \num{64}-\si{\bit} vector unit based on the version
0.5 draft of RISC-V's vector extension. Ara acts as a coprocessor
tightly coupled to Ariane, an open-source application-class RV64GC
core. Ara's microarchitecture was designed with scalability in
mind. To this end, it is composed of a set of identical lanes, each
hosting part of the system's vector register file and functional
units. The lanes communicate with each other via the \gls{VLSU} and
the \gls{SLDU}, responsible for executing instructions that touch all
the \gls{VRF} banks at once. These units arguably represent the weak
points when it comes to scalability, because they get wider with an
increasing number of lanes. Other architectures take an alternative
approach, having several narrow memory ports instead of a single wide
one. This approach does not solve the scalability problem, but just
deflects it further to the memory interconnect and cache subsystem.

We measured the performance of Ara using matrix multiplication,
convolution (both compute-bound), and AXPY (memory-bound)
double-precision kernels. For problems ``large enough,'' the
compute-bound kernels almost saturate the \glspl{FPU}, with the
measured performance of a \num{256 x 256} matrix multiplication only
\num{3}{\percent} below the theoretically achievable peak performance.

In terms of performance and power, we presented post-place-and-route
results for Ara configurations with two up to sixteen lanes in
\textsc{GlobalFoundries} 22FDX \gls{FDSOI} technology, and showed that
Ara achieves a clock frequency higher than \SI{1}{\giga\hertz} in the
typical corner. Our results indicate that our design is
$\num{2.5}\times$ more energy efficient than Ariane alone when running
an equivalent benchmark. An instance of our design with sixteen lanes
achieves up to about \SI{41}{\dpgiga\flops\per\watt} running
computationally intensive benchmarks, comparable to the energy
efficiency of the equivalent Hwacha implementation.

\review{We decided not to restrain the performance analysis to very
  large problems,} and observed a performance degradation for problems
whose size is comparable to the number of vector lanes. This is not a
limitation of Ara per se, but rather of vector processors in general,
when coupled to a single-issue in-order core. The main reason for the
low \gls{FPU} utilization for small problems is the rate at which the
scalar core issues vector instructions. With our MATMUL
implementation, Ariane issues a vector \gls{FMA} instruction every
five cycles, and the shorter the vector length is, the more vector
instructions are required to fill the pipeline. \review{By decoupling
  operand fetch and result write-back, Ara tries to eliminate bubbles
  that would have a significant impact on short-lived vector
  instructions. While the achieved performance in this case is far
  from the peak, it is nonetheless close to the instruction issue rate
  performance boundary.}

To this end, we believe that it would be interesting to investigate
whether and to what extent this performance limit could be mitigated
by leveraging a superscalar or \gls{VLIW}\kern-.1em-capable core to
drive the vector coprocessor. \review{While using multiple small cores
  to drive the vector lanes increases their individual utilization,
  maintaining an optimal energy efficiency might mean the usage of
  fewer lanes than physically available, \ie a lower overall
  utilization of the functional units. In any case, care must be taken
  to find an equilibrium between the high-performance and
  energy-efficiency requirements of the design.}


\section*{Acknowledgments}

We would like to thank Frank G\"urkaynak and Francesco Conti for the
helpful discussions and insights.


\bibliographystyle{IEEEtran.bst}
\bibliography{thesis}


\appendix

\subsection{Implementation and execution of a matrix multiplication}
\label{sec:impl-exec-matr}

Here we analyze in depth the implementation and execution of the
$n \times n$ matrix multiplication. We assume the matrices are stored
in row-major order. Our implementation uses a tiled approach working
on $t$ rows of matrix $C$ at a time. \figref{alg:matmul} presents the
matrix multiplication algorithm, working on tiles of size
$t \times n$. \review{The algorithm showcases how the \gls{ISA}
  handles scalability via strip-mined
  loops~\cite{Asanovic1998}. Line~\ref{line:matmul_setvl} is uses the
  \texttt{setvl} instruction, which sets the vector length for the
  following vector instructions, and enables the same code to be used
  for vector processors with different maximum vector length
  ${\text{VLMAX}}$.}

\begin{figure}[ht]
  \begin{algorithmic}[1]
    \STATE{$c \gets 0$;}
    \WHILE[Strip-mining loop]{$c < n$} \label{line:matmul_stripmine}
    \STATE{${\text{\textit{vl}}} \gets {\text{min}}(n - c, {\text{VLMAX}})$;} \label{line:matmul_setvl}

    \STATE{$r \gets 0$;}
    \WHILE{$r < n$}

    \FOR[Phase I]{$j \gets 0$ \TO ${\text{min}}(r, t)-1$}
    \STATE{Load row $C[r + j, c]$ into vector register $v_{C_j}$;}
    \ENDFOR

    \FOR[Phase II]{$i \gets 0$ \TO $n-1$}
    \STATE{Load row $B[i, c]$ into vector register $v_B$;}
    \FOR{$j \gets 0$ \TO ${\text{min}}(r, b)-1$} \label{line:matmul_kernel}
    \STATE{Load element $A[j,i]$;}
    \STATE{Broadcast $A[j,i]$ into vector register $v_A$;}
    \STATE{$v_{C_j} \gets v_A v_B + v_{C_j}$;}
    \ENDFOR
    \ENDFOR

    \FOR[Phase III]{$j \gets 0$ \TO ${\text{min}}(r,t)-1$}
    \STATE{Store vector register $v_{C_j}$ into $C[r + j, c]$;}
    \ENDFOR

    \STATE{$r \gets r + t$;}
    \ENDWHILE

    \STATE{$c \gets c + {\text{\textit{vl}}}$;}
    \ENDWHILE
  \end{algorithmic}
  \caption{Algorithm for the matrix multiplication $C \gets AB + C$.}
  \label{alg:matmul}
\end{figure}

Once inside the strip-mined loop, there are three distinct computation
phases:
\begin{inparaenum}[I)]
\item read a block of matrix $C$;
\item the actual computation of the matrix multiplication, and;
\item write the result to memory.
\end{inparaenum}
Phases I and III take $\BigO{(n)}$ cycles, whereas the phase II takes
$\BigO{(n^2)}$ cycles. The core part of \figref{alg:matmul} is the
\emph{for} loop of line~\ref{line:matmul_kernel}, where most of the
time is spent and where the \glspl{FPU} are used. \lstref{lst:matmul}
shows the resulting RISC-V vector assembly code for the phase II of
the matrix multiplication, considering a block size of four rows. We
ignore some control flow instructions at the start and end of
\lstref{lst:matmul}, which handle the outer \emph{for} loop.

\begin{center}
\begin{minipage}{0.9\linewidth}
  \centering
  \begin{lstlisting}[caption={Excerpt of the matrix
      multiplication in RISC-V Vector extension
      assembly, with a block size of four rows.},
    label=lst:matmul, escapechar=!]
; a0:  pointer to A
; a1:  pointer to B
; a2:  A row size
; a3:  B row size

vld   vB0, 0(a1)        ; load row of B
add   a1, a1, a3        ; bump B pointer

vld   vB1, 0(a1)        ; load row of B!\label{line:loop_matmul_pingpong}!
add   a1, a1, a3        ; bump B pointer
ld    t0,  0(a0)        ; / load element of A!\label{line:loop_matmul_begin}!
add   a0, a0, a2        ; | bump A pointer
vins  vA, t0, zero      ; | move from Ariane to Ara
vmadd vC0, vA, vB0, vC0 ; \ vector multiply-add!\label{line:loop_matmul_end}!
ld    t0, 0(a0)
add   a0, a0, a2
vins  vA, t0, zero
vmadd vC1, vA, vB1, vC1
ld    t0, 0(a0)
add   a0, a0, a2
vins  vA, t0, zero
vmadd vC2, vA, vB2, vC2
ld    t0, 0(a0)
add   a0, a0, a2
vins  vA, t0, zero
vmadd vC3, vA, vB0, vC3

vld   vB0, 0(a1)        ; load row of B!\label{line:loop_matmul_pingpong_2}!
add   a1, a1, a3        ; bump B pointer
ld    t0,  0(a0)        ; / load element of A
add   a0, a0, a2        ; | bump A pointer
vins  vA, t0, zero      ; | move from Ariane to Ara
vmadd vC0, vA, vB1, vC0 ; \ vector multiply-add
...
ld    t0,  0(a0)
add   a0, a0, a2
vins  vA, t0, zero
vmadd vC3, vA, vB1, vC3
  \end{lstlisting}
\end{minipage}
\end{center}

After loading one row of matrix $B$, the kernel consists of four
repeating instructions, responsible for, respectively:
\begin{inparaenum}[i)]
\item load the element $A[j,i]$ into a general-purpose register
  \texttt{t0};
\item bump address $A[j,i]$ preparing for next iteration;
\item broadcast scalar register \texttt{t0} into vector register $v_A$;
\item multiply-add instruction $v_{C_i} \gets v_A v_B + v_{C_i}$.
\end{inparaenum}
As Ariane is a single-issue core, this kernel runs in at least four
cycles. In steady state, however, we measure that each loop iteration
runs in five cycles. The reason for this, as shown in the pipeline
diagram of \figref{fig:execution}, is one bubble due to the data
dependence between the scalar load (which takes two cycles) and the
broadcast instruction.

\begin{figure}[ht]
  \centering
  \setlength{\tabcolsep}{3.5pt}
  \renewcommand{\arraystretch}{.7}
  \begin{tabular}{lcccccccc}
    \toprule
    \textbf{Instruction} & \multicolumn{8}{c}{\textbf{Cycle}}     \\\cmidrule{2-9}
                         & 1  & 2  & 3   & 4  & 5  & 6  & 7  & 8  \\\midrule
    \textsc{ld}          & IS & EX & EX  & CO                     \\\midrule
    \textsc{add}         &    & IS & EX  & CO                     \\\midrule
    \textsc{vins}        &    &    & --- & IS & EX & EX & CO      \\\midrule
    \textsc{vmadd}       &    &    &     &    & IS & EX & EX & CO \\\midrule
    \textsc{ld}          &    &    &     &    &    & IS & EX & EX \\\bottomrule
  \end{tabular}
  \caption{Pipeline diagram of the matrix multiplication kernel. Only
    three pipeline stages are highlighted: IS is Instruction Issue, EX
    is Execution Stage, CO is Commit Stage. Ariane has two commit
    ports into the scoreboard.}
  \label{fig:execution}
\end{figure}

\begin{figure}[ht]
  \centering
  \begin{tikzpicture}
    \begin{groupplot}[
      group style = {group size = 1 by 3, vertical sep = 1.5em},
      title style = {yshift = -1ex},
      xmajorticks = false,
      ytick = {0,50,...,100},
      xmax = 10,
      enlargelimits = false,
      grid = major,
      height = 2.5cm,
      /tikz/font=\footnotesize]

      \nextgroupplot[title = \textsc{ld}]
      \addplot [const plot, smooth, fill = color3, draw = color3] table [x expr=\thisrowno{0}*0.016, y expr=\thisrowno{1}*100] {LD};

      \nextgroupplot[title = \textsc{fpu}, ylabel = {Utilization [\si{\percent}]}]
      \addplot [const plot, smooth, fill = color2, draw = color2] table [x expr=\thisrowno{0}*0.016, y expr=\thisrowno{1}*100] {FPU};

      \nextgroupplot[title = \textsc{st}, xlabel = {Time [$\times 10^3$ cycles]}, xmajorticks = true]
      \addplot [const plot, smooth, fill = color1, draw = color1] table [x expr=\thisrowno{0}*0.016, y expr=\thisrowno{1}*100] {ST};
    \end{groupplot}
  \end{tikzpicture}
  \caption{Utilization of Ara's functional units for a \num{32 x 32}
    matrix multiplication on an Ara instance with four lanes.}
  \label{fig:matmul_utilization}
\end{figure}
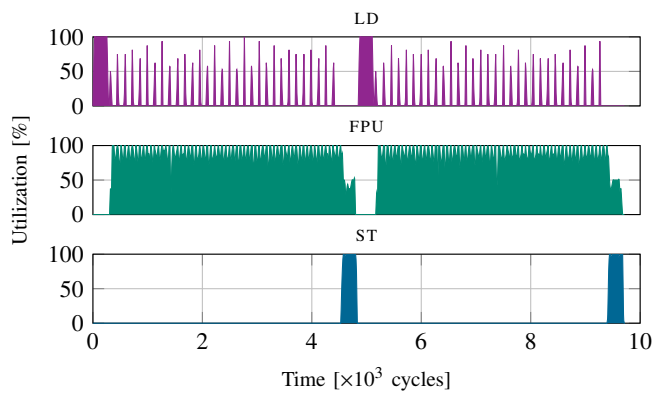

We used loop unrolling and software pipelining to code the algorithm
of \figref{alg:matmul} as our C implementation. The use of these
techniques to improve performance is visible in
\lstref{lst:matmul}. We unrolled of the \emph{for} loop of
line~\ref{fig:matmul_utilization} in \figref{alg:matmul}, which
correspond to
lines~\ref{line:loop_matmul_begin}-\ref{line:loop_matmul_end},
repeated $t$ times on the following lines in \lstref{lst:matmul}. This
avoids any branching at the end of the loop. Moreover, two vectors
hold rows of matrix $B$. This double buffering allows for the
simultaneous loading of one row in vector \texttt{vB1}, in
line~\ref{line:loop_matmul_pingpong}, while \texttt{vB0} is used for
the \glspl{FMA}, as in line~\ref{line:loop_matmul_end} in
\lstref{lst:matmul}. After line~\ref{line:loop_matmul_pingpong_2},
\texttt{vB1} is used for the computation, while another row of $B$ is
loaded into \texttt{vB0}.

The three phases of the computation can be distinguished clearly in
\figref{fig:matmul_utilization}, which shows the utilization of the
\gls{VLSU} and \gls{FPU} for a \num{32x32} matrix multiplication on a
four-lane Ara instance. Note how the \glspl{FPU} are almost fully
utilized during phase II, while being almost idle otherwise.

\begin{IEEEbiography}[{\includegraphics[width=1in,height=1.25in,clip,keepaspectratio]{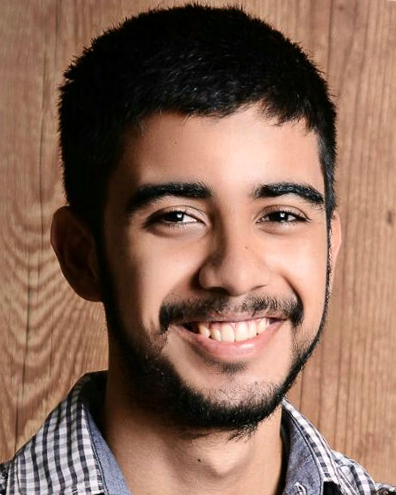}}]{Matheus Cavalcante}
  received the M.Sc.\ degree in Integrated Electronic Systems from the
  Grenoble Institute of Technology (Phelma), France, in 2018. He is
  currently pursuing a Ph.D.\ degree at the Integrated Systems
  Laboratory of ETH Z\"urich, Switzerland. His research interests
  include high performance compute architectures and interconnection
  networks.
\end{IEEEbiography}

\begin{IEEEbiography}[{\includegraphics[width=1in,height=1.25in,clip,keepaspectratio]{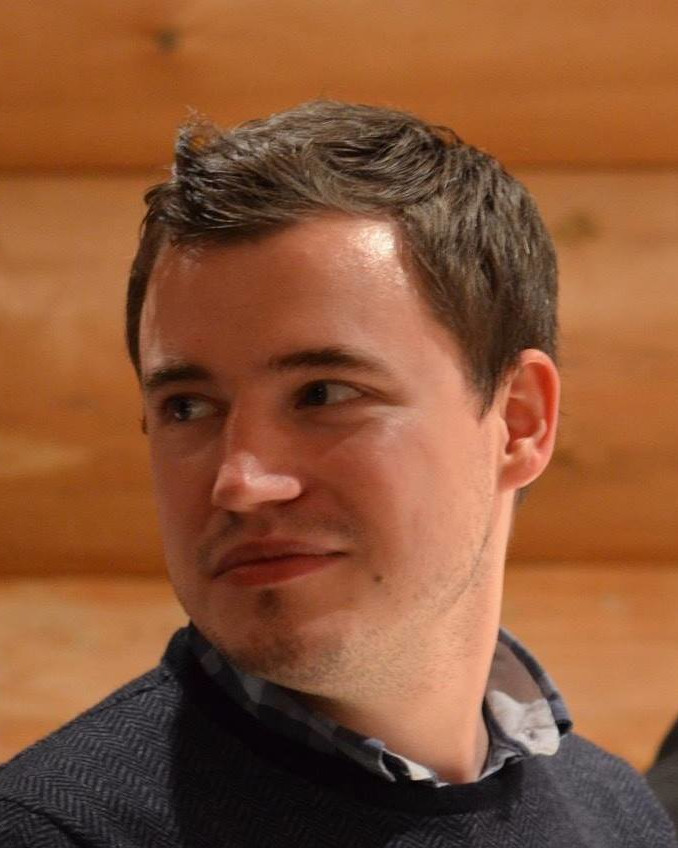}}]{Fabian Schuiki}
  received the B.Sc.\ and M.Sc.\ degree in electrical engineering from
  the ETH Z\"urich in \num{2014} and \num{2016}, respectively. He is
  currently pursuing a Ph.D.\ degree with the Digital Circuits and
  Systems group of Luca~Benini. His research interests include
  transprecision computing as well as near- and in-memory processing.
\end{IEEEbiography}

\begin{IEEEbiography}[{\includegraphics[width=1in,height=1.25in,clip,keepaspectratio]{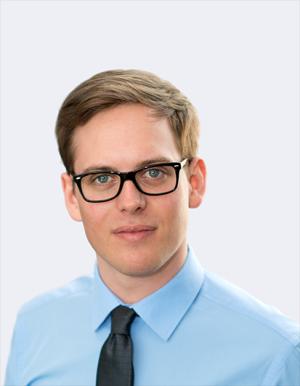}}]{Florian Zaruba}
  received his B.Sc.\ degree from TU Wien in 2014 and his M.Sc.\ from
  the ETH Z\"urich in \num{2017}. He is currently pursuing a Ph.D.\
  degree at the Integrated Systems Laboratory. His research interests
  include design of very large scale integration circuits and high
  performance computer architectures.
\end{IEEEbiography}

\begin{IEEEbiography}[{\includegraphics[width=1in,height=1.25in,clip,keepaspectratio]{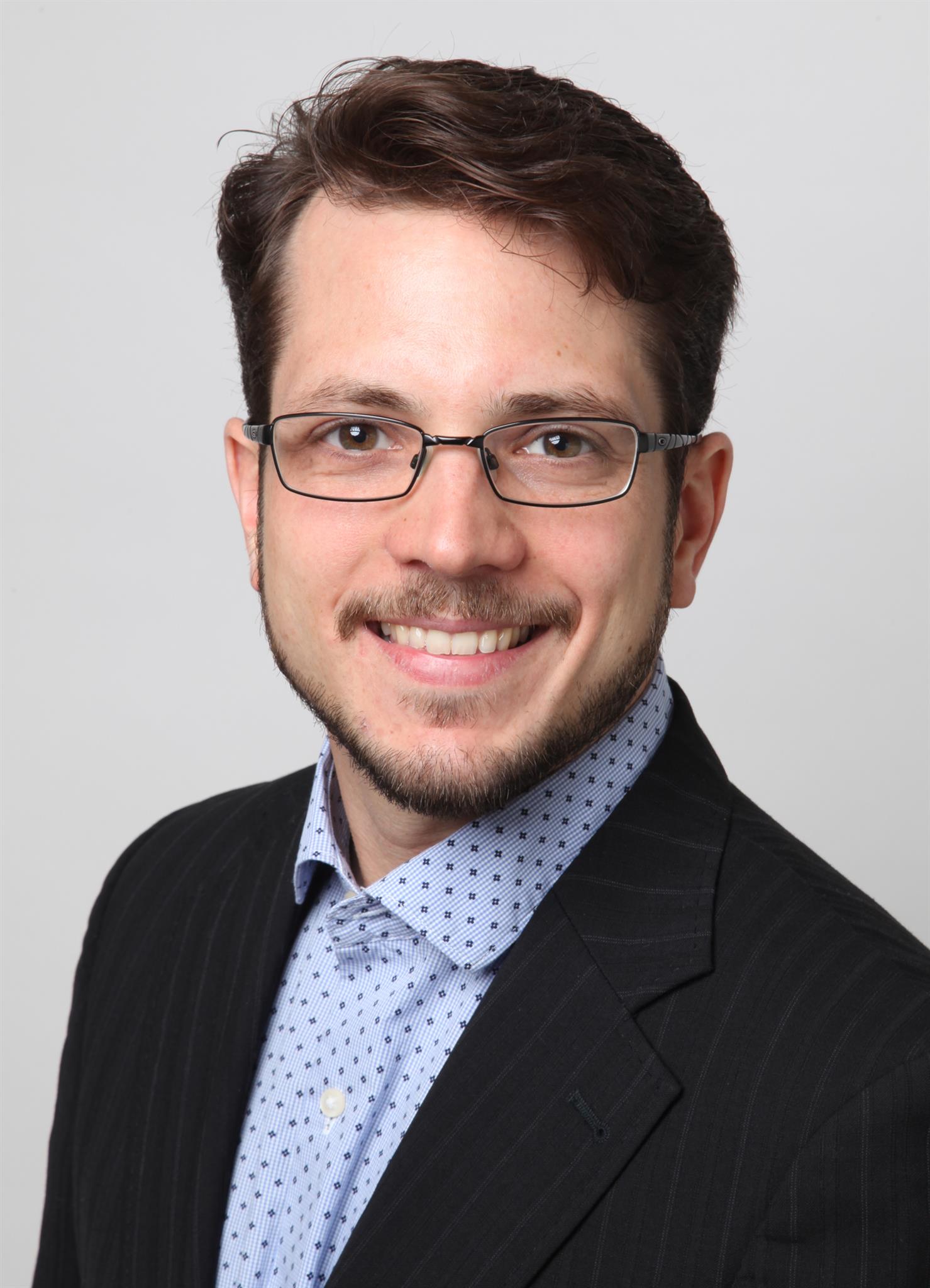}}]{Michael Schaffner}
  received his M.Sc.\ and Ph.D.\ degrees from ETH Z\"urich,
  Switzerland, in \num{2012} and \num{2017}. He has been a research
  assistant at the Integrated Systems Laboratory, ETH Z\"urich, and
  Disney Research, Z\"urich, from \num{2012} to \num{2017}, where he
  was working on digital signal and video processing. From \num{2017}
  to \num{2018} he has been a postdoctoral researcher at the
  Integrated Systems Laboratory, ETH Z\"urich, focusing on the design
  of RISC-V processors and efficient co-processors. Since \num{2019},
  he has been with the ASIC development team at Google Cloud
  Platforms, Sunnyvale, USA, where he is involved in processor
  design. Michael Schaffner received the ETH Medal for his Diploma
  thesis in \num{2013}.
\end{IEEEbiography}

\begin{IEEEbiography}[{\includegraphics[width=1in,height=1.25in,clip,keepaspectratio]{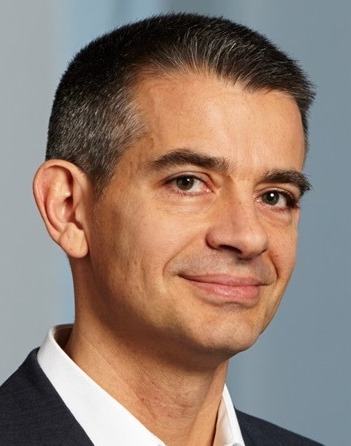}}]{Luca Benini}
  holds the chair of digital Circuits and systems at ETH Z\"urich and
  is Full Professor at the Universit\`a di Bologna. In
  \num{2009}-\num{2012} he served as chief architect in
  STMicroelectronics France. Dr.\ Benini's research interests are in
  energy-efficient computing systems design, from embedded to
  high-performance. He is also active in the design ultra-low power
  VLSI Circuits and smart sensing micro-systems. He has published more
  than \num{1000} peer-reviewed papers and five books. He is a Fellow
  of the ACM and a member of the Academia Europaea. He is the
  recipient of the \num{2016} IEEE CAS Mac Van Valkenburg award and of
  the \num{2019} IEEE TCAD Donald O.\ Pederson Best Paper Award.
\end{IEEEbiography}

\end{document}